\documentclass[10pt,conference]{IEEEtran}
\usepackage{cite}
\usepackage{amsmath,amssymb,amsfonts}
\usepackage{algorithmic}
\usepackage{algorithm}
\usepackage{graphicx}
\usepackage{textcomp}
\usepackage{xcolor}
\usepackage[hyphens]{url}
\usepackage{fancyhdr}
\usepackage{hyperref}
\usepackage{xspace}
\usepackage{booktabs}
\usepackage{multirow}

\usepackage{tikz}
\usepackage{color,soul}

\newcommand{\name}{\textit{Prosperity}\xspace}

\newcommand{\revision}[1]{{#1}}

\newcommand{\Fig}[1]{Fig.~\ref{#1}}

\newcommand{\Tbl}[1]{Tbl.~\ref{#1}}
\newcommand{\Sec}[1]{Sec.~\ref{#1}}
\newcommand{\Reb}[1]{\textcolor{red}}

\definecolor{mycolorblue}{RGB}{81,140,230}
\definecolor{mycolorgreen}{RGB}{166,213,95}
\definecolor{mycoloryellow}{RGB}{242,210,98}
\definecolor{mycolorpink}{RGB}{242,96,119}
\definecolor{mycolorpurple}{RGB}{149,90,189}
\definecolor{mycolororange}{RGB}{255,199,7}

\newcommand\circlednumberpink[1]{%
  \begin{tikzpicture}[baseline=(char.base)]
    \node[shape=circle,,fill=mycolorpink,inner sep=1pt] (char) {\textcolor{white}{\scriptsize\sffamily\bfseries#1}};
\end{tikzpicture}}

\newcommand\circlednumberpurple[1]{%
  \begin{tikzpicture}[baseline=(char.base)]
    \node[shape=circle,,fill=mycolorpurple,inner sep=1pt] (char) {\textcolor{white}{\scriptsize\sffamily\bfseries#1}};
\end{tikzpicture}}

\newcommand\circlednumberorange[1]{%
  \begin{tikzpicture}[baseline=(char.base)]
    \node[shape=circle,,fill=mycolororange,inner sep=1pt] (char) {\textcolor{black}{\scriptsize\sffamily\bfseries#1}};
\end{tikzpicture}}

\newcommand\circlednumberorangetiny[1]{%
    \begin{tikzpicture}[baseline=(char.base)]
      \node[shape=circle,,fill=mycolororange,inner sep=0pt] (char) {\textcolor{black}{\scriptsize\sffamily\bfseries#1}};
    \end{tikzpicture}}

\usepackage{wrapfig}

\newcommand\blankfootnote[1]{%
            \let\thefootnote\relax\footnotetext{#1}%
            \let\thefootnote\svthefootnote%
          }

\newcommand{\hpcayear}{2025}

\title{\name: Accelerating Spiking Neural Networks via Product Sparsity
}

\def\hpcacameraready{} 

\newcommand\hpcaauthors{Chiyue Wei$\dagger$, Cong Guo$\dagger$\IEEEauthorrefmark{1}, Feng Cheng$\dagger$, Shiyu Li$\dagger$, Hao (Frank) Yang$\ddagger$, Hai (Helen) Li$\dagger$, Yiran Chen$\dagger$}
\newcommand\hpcaaffiliation{Duke University$\dagger$, Johns Hopkins University$\ddagger$}
\newcommand\hpcaemail{\{chiyue.wei, cong.guo, feng.cheng, shiyu.li, hai.li, yiran.chen\}@duke.edu, haofrankyang@jhu.edu}

\def\aeopen{}           
\def\aereviewed{}     
\def\aereproduced{} 

\author{
  \ifdefined\hpcacameraready
    \IEEEauthorblockN{\hpcaauthors{}}
      \IEEEauthorblockA{
        \hpcaaffiliation{} \\
        \hpcaemail{}\\
      }
  \else
    \IEEEauthorblockN{\normalsize{HPCA \hpcayear{} Submission
      \textbf{\#\hpcasubmissionnumber{}}} \\
      \IEEEauthorblockA{
        Confidential Draft \\
        Do NOT Distribute!!
      }
    }
  \fi 
}

\fancypagestyle{camerareadyfirstpage}{%
  \fancyhead{}
  
  \fancyhead[C]{
    \ifdefined\aeopen
    \parbox[][12mm][t]{13.5cm}{\hpcayear{} IEEE International Symposium on High-Performance Computer Architecture (HPCA)}    
    \else
      \ifdefined\aereviewed
      \parbox[][12mm][t]{13.5cm}{\hpcayear{} IEEE International Symposium on High-Performance Computer Architecture (HPCA)}
      \else
      \ifdefined\aereproduced
      \parbox[][12mm][t]{13.5cm}{\hpcayear{} IEEE International Symposium on High-Performance Computer Architecture (HPCA)}
      \else
      \parbox[][0mm][t]{13.5cm}{\hpcayear{} IEEE International Symposium on High-Performance Computer Architecture (HPCA)}
    \fi 
    \fi 
    \fi 
    \ifdefined\aeopen 
      \includegraphics[width=12mm,height=12mm]{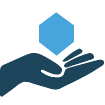}
    \fi 
    \ifdefined\aereviewed
      \includegraphics[width=12mm,height=12mm]{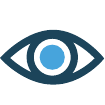}
    \fi 
    \ifdefined\aereproduced
      \includegraphics[width=12mm,height=12mm]{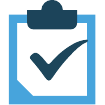}
    \fi
  }
  \fancyfoot[C]{}
}
\begin{document}
\maketitle

\ifdefined\hpcacameraready 
  \thispagestyle{camerareadyfirstpage}
  \pagestyle{empty}
\else
  \thispagestyle{plain}
  \pagestyle{plain}
\fi

\newcommand{\hpcaheight}{0mm}
\ifdefined\eaopen
\renewcommand{\hpcaheight}{12mm}
\fi


\begin{abstract}
Spiking Neural Networks (SNNs) are highly efficient due to their spike-based activation, which inherently produces bit-sparse computation patterns.
Existing hardware implementations of SNNs leverage this sparsity pattern to avoid wasteful zero-value computations, yet this approach fails to fully capitalize on the potential efficiency of SNNs.
This study introduces a novel sparsity paradigm called \textit{Product Sparsity}, which leverages combinatorial similarities within matrix multiplication operations to reuse the inner product result and reduce redundant computations. 
{Product Sparsity} significantly enhances sparsity in SNNs without compromising the original computation results compared to traditional bit sparsity methods. 
For instance, in the \revision{SpikeBERT SNN model}, {Product Sparsity} achieves a density of only \revision{$1.23\%$} and reduces computation by \revision{$11\times$}, compared to bit sparsity, which has a density of \revision{$13.19\%$}. 
To efficiently implement {Product Sparsity}, we propose \name, an architecture that addresses the challenges of identifying and eliminating redundant computations in real-time. 
Compared to prior SNN accelerator PTB and the A100 GPU, \name achieves an average speedup of $7.4\times$ and $1.8\times$, respectively, along with energy efficiency improvements of $8.0\times$ and $193\times$, respectively. The code for \name is available at \href{https://github.com/dubcyfor3/Prosperity}{https://github.com/dubcyfor3/Prosperity}.
\end{abstract}

\section{Introduction}\label{sec:intro}

\blankfootnote{*Cong Guo is the corresponding author of this paper.}
Spiking Neural Networks (SNNs)\cite{ghosh2009spiking,maass1997networks} are brain-inspired neural network models that have gained significant attention in recent years. 
Extensive research efforts in SNN algorithm design have yielded performance comparable to Deep Neural Networks (DNNs) across various tasks, including computer vision~\cite{cao2015spiking,fang2021deep} and natural language processing~\cite{bal2024spikingbert,lv2023spikebert}.
Beyond traditional AI applications, SNNs show promise in rapidly solving optimization problems~\cite{mniszewski2019graph} and heuristically approaching NP-hard problems~\cite{jonke2016solving}, further underscoring their versatility and potential impact across diverse computational domains.

\begin{figure}
    \vspace{-10pt}
    \centering
    \includegraphics[width=0.48\textwidth]{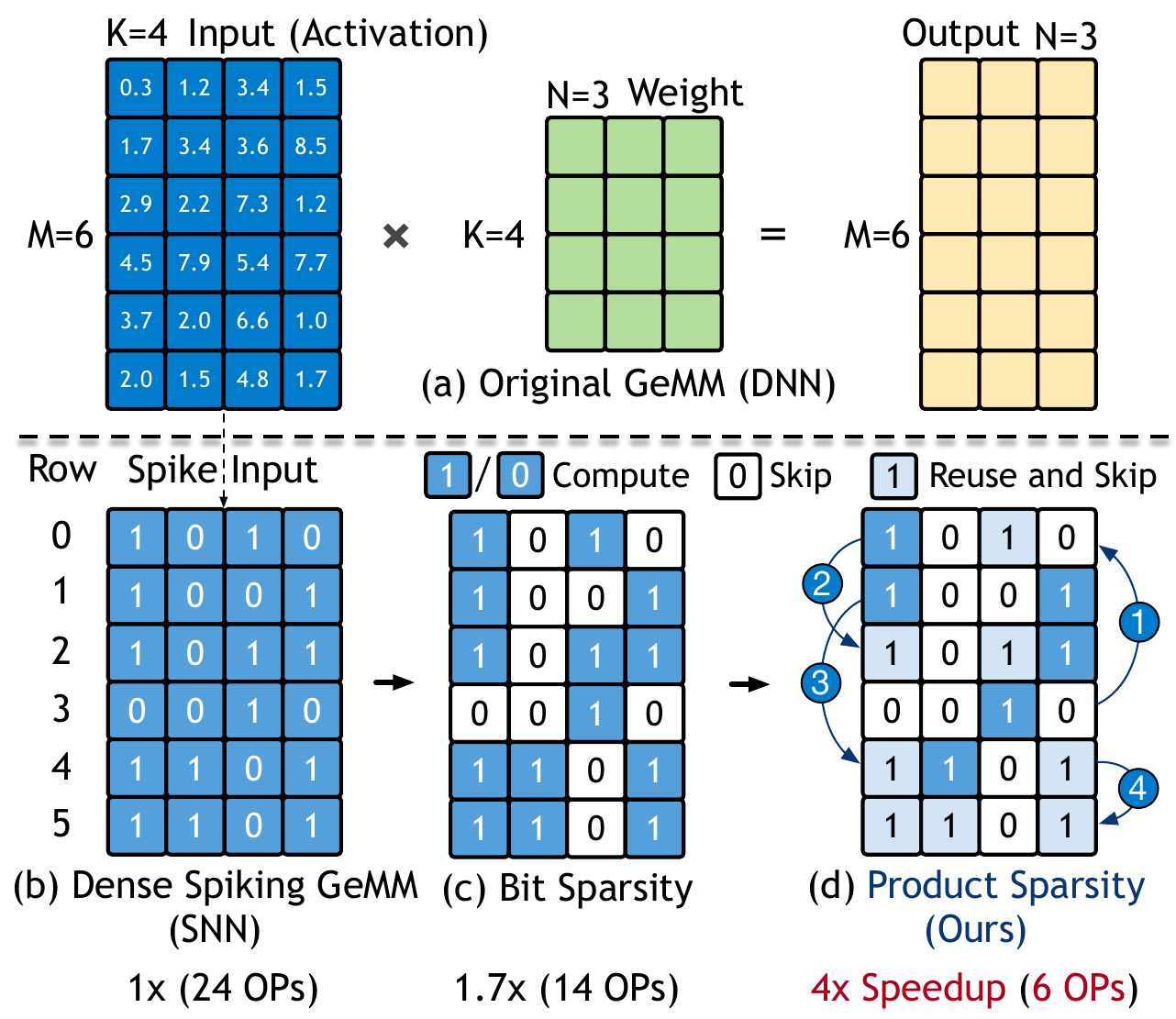}
    \vspace{-8pt}

    \caption{Comparison of DNN, SNN, Bit Sparsity, and Product Sparsity. 
    }
    \label{fig:intro}
    \vspace{-10pt}
\end{figure}

Compared to the traditional DNN models, as shown in \Fig{fig:intro}~(a), the SNN's neurons only react to information encoded as binary spikes (1 for spike and 0 for non-spike), as depicted \Fig{fig:intro}~(b), leading to sparse activations (i.e., \textbf{\textit{bit sparsity}}).
This design leads to significant computational and energy efficiency gains~\cite{pfeiffer2018deep}, which arise from the spike-driven sparse processing in SNNs, depicted in \Fig{fig:intro}~(c). 
Consequently, many SNN-specific architecture designs~\cite{lee2022parallel,narayanan2020spinalflow} leverage this \textit{bit sparsity} to enhance performance and energy efficiency, establishing a paradigm for bit sparsity-based accelerators. 
Nevertheless, our study observed that \textit{bit sparsity} alone does not fully harness the potential sparsity within SNNs.

In this study, a novel sparsity paradigm, \textbf{\textit{Product Sparsity (ProSparsity)}}, is proposed, which can reveal additional sparsity opportunities within SNN models, as illustrated in \Fig{fig:intro}~(d). 
ProSparsity is founded on the combinatorial similarity within the inner product operations of matrix multiplication. 
Specifically, in SNNs, multiple rows of the binary spike matrix usually contain a common binary sub-combination, resulting in identical inner product results when multiplied by the shared weight matrix. 
By computing the inner product of the sub-combination only once and reusing the result for all rows containing the same sub-combination, ProSparsity effectively eliminates the repeat computations of spikes within the common sub-combination. 
Consequently, ProSparsity significantly enhances sparsity and reduces redundant computations compared to traditional bit sparsity.

For example, in \Fig{fig:intro}~(d), Row 1 (1001) and Row 4 (1101) share the common sub-combination 1001. 
Thus, the inner product result of Row 1 (1001) can be reused for Row 4, with additional computation needed only for the remaining sub-combination 0100, saving two operations.
Moreover, since Row 5 is identical to Row 4, all computations for Row 5 can be skipped, and the result of Row 4 can be directly applied to Row~5. 
Consequently, ProSparsity enables significant computation skipping and result reuse. 
Our experiments demonstrate that, in the \revision{SpikeBERT SNN model~\cite{lv2023spikebert}}, ProSparsity achieves a \revision{$\mathbf{11\times}$} reduction in computation. 
In comparison, while bit sparsity achieves a density of \revision{$\mathbf{13.19\%}$}, ProSparsity reduces the density to just $\mathbf{1.23\%}$. 

To the best of our knowledge, this is the first work to accelerate SNNs with product sparsity, thereby leading to several significant challenges in designing a ProSparsity-based architecture. 
Identifying common sub-combinations presents both spatial and temporal challenges.
Firstly, for spatial relationships, ProSparsity must identify the combinatorial similarity between multiple rows. 
The complexity for identifying spatial relations among $n$ rows within an $m$-row scope is $O(m^n)$, which is not feasible at runtime. Therefore, an efficient method is needed to identify ProSparsity relations with manageable overhead.
Secondly, temporal relationships also significantly impact the efficiency of ProSparsity. 
For instance, in \Fig{fig:intro}~(d), if Row 0 (1010) is processed first, it cannot reuse the result from Row 3 (0010).
Improper execution orders will significantly undermine the effectiveness of ProSparsity. 
Thus, optimizing the execution order of spike rows is crucial for maximizing reusability and achieving high sparsity. 
Moreover, since spike activations are generated dynamically, the identification of sub-combinations and the elimination of redundant computations have to be done at runtime, which further complicates the design of ProSparsity.

To address these challenges, we propose \textbf{\textit{Prosperity}}, an architecture designed to efficiently process SNNs using product sparsity. 
Based on our analysis, we first introduce a heuristic optimization to reduce the time and space complexity of ProSparsity processing to a linear level, addressing both spatial and temporal challenges. 
We then design an efficient architecture and pipeline scheduling mechanism that overlaps the identification and elimination of ProSparsity with matrix computation, resulting in overhead-free sparse processing.
Furthermore, since ProSparsity is algorithm-agnostic, \name can support a wide range of SNNs, including emerging spiking networks such as spiking transformers~\cite{zhou2022spikformer,bal2024spikingbert}, which are not supported by existing accelerators~\cite{lee2022parallel,mao2024stellar,liu2022sato}.

Compared to state-of-the-art SNN accelerators like PTB~\cite{lee2022parallel} and the A100 GPU~\cite{nvidia_a100}, \name achieves an average speedup of $7.4\times$ and $1.8\times$, respectively, along with energy efficiency improvements of $8.0\times$ and $193\times$, respectively.
Our contributions are summarized as follows:

\begin{itemize}
    \item We propose a novel sparsity paradigm, \textbf{\textit{Product Sparsity}}, that utilizes the combinatorial similarity feature in SNN activations to significantly enhance sparsity and reduce computations in SNNs.
    \item To tackle high complexity and dependency challenges in ProSparsity, we present a dedicated architecture, \textbf{\name}, that efficiently identifies and utilizes ProSparsity within linear complexity. 
    \item We design an efficient pipeline scheduling method to further overlap the ProSparsity processing with the matrix computation, achieving overhead-free sparse processing.
    \item Finally, \name architecture effectively accelerates a wide range of SNNs, achieving significant speedup and energy efficiency over baselines.
\end{itemize}
\section{Background}

\subsection{Spiking Neural Networks} \label{sec:SNN}

Spiking Neural Networks (SNNs)~\cite{ghosh2009spiking,ponulak2011introduction}, said to be the third generation of neural networks~\cite{maass1997networks}, mimic the behavior of biological brain neural networks.
SNNs have a similar structure as traditional Deep Neural Networks (DNNs)~\cite{lecun2015deep,goodfellow2016deep} but differ in two aspects: 1) the information (activation) propagating in the network is a binary spike event in a time sequence instead of a floating point number. 2) the activation function is replaced by the membrane potential (MP) update~\cite{gerstner2014neuronal} and spike fire behavior of the spiking neuron. 
A single forward pass of SNN is accomplished in a certain number of time steps.
In each time step, a spiking neuron receives spikes from neurons in the previous layer and then integrates received spikes according to the corresponding weight to get input current; it then uses the input current to update its MP and send a spike to neurons in the next layer if its MP exceeds a certain threshold.
There are various spiking neuron model~\cite{gerstner2014neuronal,gerstner1995time,hodgkin1952quantitative,izhikevich2003simple}.
We consider the most widely used leaky integrate and fire (LIF) model~\cite{gerstner2014neuronal}.

SNNs are more energy efficient than DNNs due to their binary activation.
In DNN, the major computation is the matrix multiplication between floating point activation and weight matrix, i.e., general matrix multiplication (GeMM).
In SNN, the activation of an SNN layer consists of a set of binary spike matrices, each representing a time step of SNN and sharing the same weight.
Therefore, we can unroll and concatenate all spike matrices in different time steps to get a single binary spike matrix for an SNN layer.
The major computation of the SNN model is the matrix multiplication of the binary spike matrix and floating point weight matrix~\cite{deng2020rethinking,pfeiffer2018deep}.
We define this unique operation in SNNs as \textbf{Spiking GeMM}.
Our experiments show that more than 98\% percent of operations in SNNs are spiking GeMM.
In spiking GeMM, the binary spike matrix only consists of 0s and 1s, i.e., \textbf{\textit{bit sparsity (BitSparsity)}}.
Therefore, the computation becomes a sparse addition. If an element is 1, then the corresponding weight is straightly accumulated to the result; if an element is 0, then the corresponding weight is skipped.
Prior works~\cite{narayanan2020spinalflow,lee2022parallel,park2020t2fsnn,liu2022randomize} leveraged the sparse and multiplication-free feature in spiking GeMM to accelerate SNN. 

\subsection{Spiking CNNs and Spiking Transformers}

The most widely used network structure in SNNs is convolutions, which is spiking CNNs~\cite{sengupta2019going,fang2021deep}. 
Besides, SNNs in transformer architecture have developed very recently while demonstrating superior performance than spiking CNNs on various tasks~\cite{zhou2022spikformer,yao2024sdt,lv2023spikebert,bal2024spikingbert,wang2023masked}.

The spiking CNNs and spiking transformers differ from their DNN counterparts in the two aspects mentioned in \Sec{sec:SNN}.
The key operation in spiking CNNs and spiking transformers is spiking GeMM. 
In spiking CNNs, the convolution is performed between a spike input feature map in multiple time steps and a weighted kernel, generating an output feature map in multiple time steps. 
This operation can be translated to spiking GeMM by im2col~\cite{chellapilla2006high}.

In spiking transformers, the linear projection layers (query, key, value, output projection, and feed-forward layer) are naturally spiking GeMM, where spike matrix of shape $(T\times L,d_i)$ is multiplied with weight matrix of shape $(d_i, d_o)$ and get output matrix of shape $(T\times L, d_o)$, where $T, L, d_i, d_o$ denotes time steps, sequence length, input dimension, and output dimension.
Different spiking transformer models incorporate diverse spiking attention blocks that are distinct from linear layers. Consequently, operations in spiking attention are not efficiently supported by existing SNN ASICs~\cite{narayanan2020spinalflow,lee2022parallel,mao2024stellar} or neuromorphic chips~\cite{davies2018loihi,deng2020tianjic}.

\subsection{SNN Accelerators} \label{sec:snn_accelerator}

Previous SNN accelerators have been exploring utilizing the bit sparsity of SNNs for efficient inference, as summarized in \Tbl{tab:prelim}.
PTB~\cite{lee2022parallel} is a systolic-array-based~\cite{kung1982systolic} architecture that parallelly processes spikes in a structured sparsity manner.
It employs time batching~\cite{narayanan2020spinalflow}, and further group spike information within a time window. If a spike occurs in a group, all time steps within that window are processed, while groups without spikes are squeezed out to enhance utilization of systolic array for sparse input.

Stellar~\cite{mao2024stellar} is an algorithm-hardware co-design accelerator. 
It is also a systolic-array-based architecture that process spikes in a structured sparsity manner. 
Stellar proposes using FS neuron~\cite{stockl2021optimized} model in the SNNs; this FS neuron is featured to have fewer spikes than the LIF neuron,  achieving higher spike sparsity. 
The improvement of sparsity is caused by algorithmic modification, which limits the application scenarios of the accelerator.


\section{Product Sparsity} \label{sec:product sparsity}
In this section, we first introduce the product sparsity in spiking GeMM.
Then, we describe some basic definitions and propose an efficient optimization for product sparsity, significantly reducing its complexity.

\begin{table}[t]
    \centering
    \vspace{-10pt}
    \caption{Comparsion with previous work on VGG-16.}
    \vspace{-8pt}
    \resizebox{0.49\textwidth}{!}{
    \begin{tabular}{l|c|c|c|c}
    \toprule
        Study & Dense & PTB~\cite{lee2022parallel} & Stellar~\cite{mao2024stellar} & \textbf{\name} \\  \midrule
        \multirow{ 2}{*}{Feature} & \multirow{ 2}{*}{-} & Long & Specific  & \textbf{Algorithm-} \\ 
        &  & Steps & Neuron & \textbf{agnostic} \\ \midrule
        Sparsity & \multirow{2}{*}{None} & Structured & Structured & \textbf{Unstructured} \\
        Pattern & ~ & BitSparsity & BitSparsity & \textbf{ProSparsity} \\
        \midrule
        Bit Density & 100\% & 34.21\% & 9.80\% & 34.21\% \\\midrule
        Pro Density & 100\% &  - & - & \textbf{2.79\%} \\\midrule
        Speedup & 1.00$\times$ & 1.86$\times$ & 5.97$\times$ & \textbf{17.55$\times$} \\
        \bottomrule
    \end{tabular}
    }
    \label{tab:prelim}
    \vspace{-10pt}
\end{table}

\subsection{Definition} \label{sec:PS motivation}
\begin{figure}
    \centering
    \includegraphics[width=0.47\textwidth]{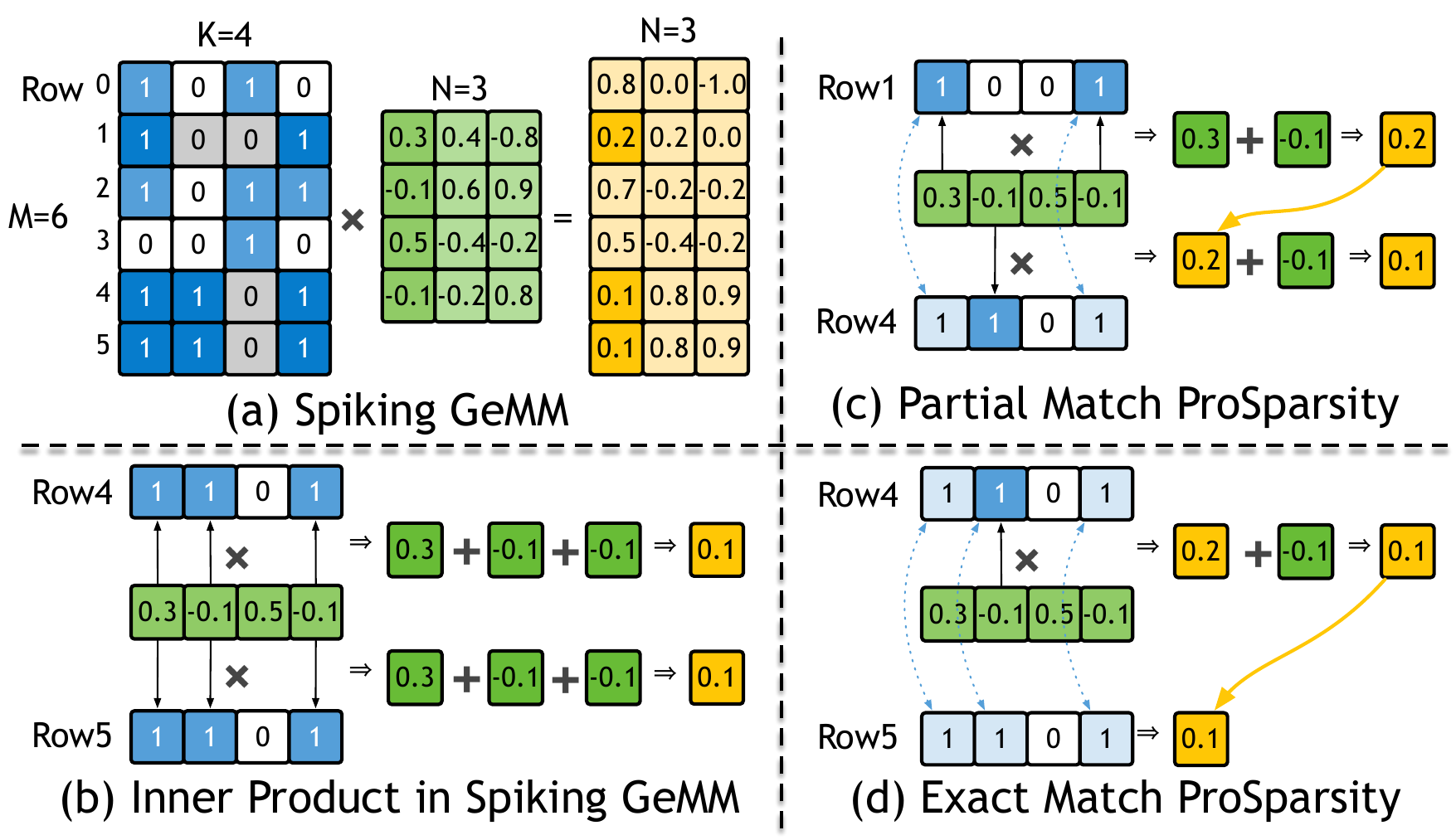}
    \vspace{-8pt}
    \caption{An example of spiking GeMM and product sparsity}
    \label{fig:spikingGeMM}
    \vspace{-10pt}
\end{figure}

In the spiking GeMM, the binary nature of the spike matrix inherently introduces bit sparsity, i.e., the zero-value operand can be skipped during computation.
However, there are more redundant computations beyond the zero-value operand.
To demonstrate this redundancy, we illustrate a spiking GeMM in \Fig{fig:spikingGeMM}~(a). 
For a $M \times K$ spike matrix, a $K \times N$ weight matrix, and a $M \times N$ output matrix, 
the spiking GeMM can be computed by multiple BitSparsity-based inner product operations, shown in \Fig{fig:spikingGeMM}~(b).
Due to the binary value of the spike matrix, the inner product is equivalent to the accumulation of multiple weight values, which are ``selected'' by the 1-values in the corresponding row of the spike matrix.
Based on the arithmetic rule of GeMM, the $i$-th column of the spike matrix will ``select'' the identical $i$-th row of the weight matrix in the inner product operations.
\revision{Two rows in the spike matrix contain a common binary sub-combination if they share identical 1-value positions, either partially or fully.}
Consequently, the common binary sub-combination will generate the same inner product result when computing the same column of the output matrix.
For example, in \Fig{fig:spikingGeMM}~(b), Row~4 and Row~5 have the same result because they have the identical binary combination (1101).
Obviously, we can compute the common binary sub-combination once and reuse its result for all rows containing this sub-combination, thereby significantly reducing computational redundancy in spiking GeMM.
We call this paradigm \textbf{\textit{product sparsity (ProSparsity)}}, which can skip computation and reuse the result based on the combinatorial similarities of inner product operation.

As shown in \Tbl{tab:prelim}, ProSparsity completely differs from the previous approaches, which depend on the algorithm and may impact the accuracy.
ProSparsity is an algorithm-agnostic and lossless method for the spiking GeMM.
Based on the potential combinatorial similarities in the spike matrix, ProSparsity can significantly reduce computations. For the VGG-16 SNN model, ProSparsity can save more than $18\times$ computations and achieve $9.4\times$ speedup over the BitSparsity of PTB~\cite{lee2022parallel}.

Despite ProSparsity providing a simple yet effective way to identify the redundancy in SNN, its implementation is a nontrivial task, as mentioned in \Sec{sec:intro}.
There are two main challenges in processing ProSparsity: Efficiently identifying spatial and temporal relationships.
We will introduce each challenge and provide an efficient heuristic solution to address these challenges.

\subsection{Spatial Relationship} \label{sec:relation}
For ProSparsity, it will cause significant complexity to identify the combinatorial similarities.
Assuming we have $m$ rows and aim to identify the common sub-combination for $n$ rows, the resulting complexity will be $O(m^n)$.
For example, if we want to find the common sub-combination between only two rows within $m$ row, we need to match $\frac{m \times (m-1)}{2}$ times to finish all identification for every two rows.
For a better illustration, we use a set to represent the spike row. 
We define $S_i$ as the spike set of the spike row $i$; the spike set consists of the column index where a spike exists \[ S_i = \{j| M[i,j] = 1, j \in \text{Column IDs}\}, \] where $M$ is the spike matrix. For example, Row~0 (1010) in \Fig{fig:spikingGeMM}~(a) can be represented by $S_0 = \{0, 2\}$.
\revision{The identification process is to inspect the set relationship among the $n$ sets represented by $n$ rows, called spatial relationship.}
Basically, when $n\ge 3$, the complexity of the identification will be unacceptable.
Therefore, in this study, we only consider the two-row spatial relationship, which can achieve high enough sparsity for SNNs, with the complexity $O(m^2)$.

Basically, we discover 3 types of ProSparsity relationship when two sets are intersected, $A = S_i \cap S_j, A\ne \phi, i \ne j$.
We will define 3 relationships with the intersection set $A$.

\textbf{Partial Match (PM).}
We define the Partial Match (PM) relationship for two spike rows as follows: 
\[ A=S_j, A\ne S_i. \]
This indicates that $S_j$ is the proper subset of $S_i$.
For example, in \Fig{fig:spikingGeMM}~(c), Row~1 (1001) is the proper subset of Row~4 (1101).
Consequently, the spike row $S_4$ (Row~4) can reuse the inner product result (0.2) of $S_1$ (Row~1).
After that, for $S_4$, we can skip the spikes in $S_1$ and accumulate the rest weight values (-0.1) that correspond to spikes in $S_4 - S_1$ (0100).

\textbf{Exact Match (EM).}
Exact Match product sparsity means two rows are identical, i.e., $$A = S_i = S_j.$$
As exemplified in \Fig{fig:spikingGeMM}~(d), we have computed the inner product result (0.1) of Row~4 (1101); we can reuse it as the result of Row~5 (1101) and skip all the spikes without any computation.

\textbf{Intersection.}
Finally, when $A \ne S_i$ and $ A\ne  S_j$, that means the two rows have an intersection relationship.
However, leveraging this scenario requires creating a new row $A$ and compute the result of $A$ first.
This will significantly increase the complexity of the architecture design. 
Hence, this study will only consider the former two relationships, EM and PM.
 
\subsection{Temporal Relationship}
\label{subsec:tr}
The temporal relationships can significantly impact efficiency. 
For instance, in \Fig{fig:spikingGeMM}~(a), if Row~0 (1010) is processed first, it cannot reuse the result from Row~3 (0010), which significantly undermines the effectiveness of product sparsity. 
Therefore, optimizing the execution order of spike rows is crucial for maximizing sparsity.

We first introduce the \textbf{\textit{Prefix}} and \textbf{\textit{Suffix}} for the temporal relationship of ProSparsity.
For the two rows with EM,  we can define the row with a smaller index as the Prefix, i.e., Row~$i$ is the Prefix of Row~$j$, when $i <j$, because the smaller index is always accessed first.
In contrast, PM relationships have a strict partial ordering, which leads to strict temporal ordering.
The Prefix in the PM relationship is the spike row with fewer elements (ones), which is the proper subset for the other row, i.e., $S_j$ is the Prefix of $S_i$, when $A=S_j, A\ne S_i$.
Correspondingly, the $S_i$ is the Suffix.

After defining spatial and temporal relationships, we can build a ProSparsity graph for the spiking matrix in the next subsection.

\subsection{Efficient Design Space Exploration} \label{sec:eff_dse}
\begin{figure}
    \centering
    \includegraphics[width=0.45\textwidth]{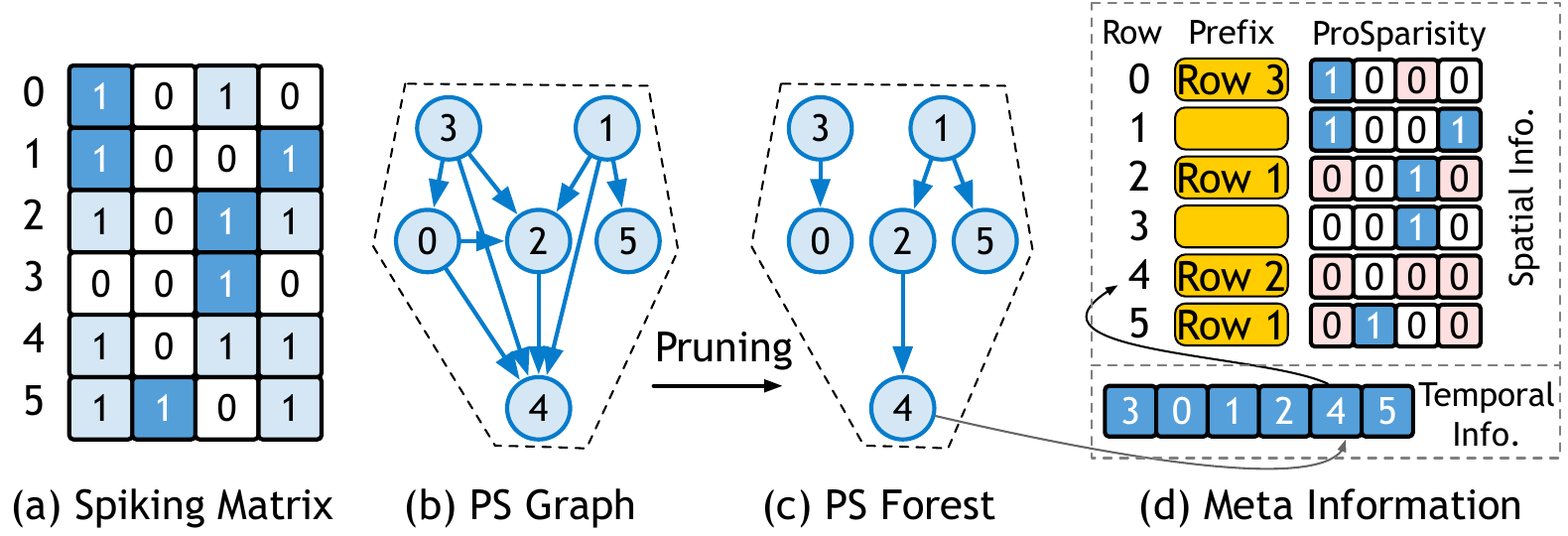}
    \vspace{-8pt}
    \caption{Efficient ProSparsity data structure.}
    \label{fig:graph_tree}
    \vspace{-10pt}
\end{figure}

Based on the spatial and temporal relationship, the ProSparsity can be constructed as a directed graph, as shown in \Fig{fig:graph_tree}.
However, the graph structure is still too complicated for the ProSparsity architecture design.
Thus, we propose a heuristic method, reducing the time and space complexity from $O(m^2)$ to $O(m)$.

\textbf{ProSparsity Graph.} 
As shown in \Fig{fig:graph_tree}~(a) and (b), each spike row in the graph is represented by a node, i.e., a spike row.
Each directed edge represents the Prefix-based partial ordering relationship, including the EM and PM edges.
The direction of the edge represents a prefix node (spike row) points to a suffix node.
Graph-based ProSparsity has both the space and time complexity with $O(m^2)$.
However, because a node may have multiple prefix nodes, we still need an extra operation to identify whether the various prefixes are available to reuse together, i.e., whether the multiple prefix nodes are disjoint.
That will significantly complicate the ProSparsity design.
Fortunately, our preliminary results show we can reduce the graph to a tree with a marginal sparsity drop.

\textbf{Heuristic Space Pruning.}
We conduct a characteristic study to explore more efficient design space for ProSparsity.
There will be a large overhead if we try to utilize two or more prefix nodes.
According to preliminary evaluation, our architecture will suffer a 30\% performance drop and larger area overhead to accommodate the second Prefix.
Moreover, \Tbl{tbl:prefix} shows the SNN density results after ProSparsity with only one prefix and two prefix nodes. 
Clearly, the first prefix node contributes to the most computation reduction compared to the second prefix nodes. 
Moreover, most nodes have only one Prefix, and due to the disjoint limitation of the prefix nodes, less than 6\% of nodes can employ the second Prefix.
This illustrates that using one Prefix for each node is capable of eliminating most of the redundancy.
Therefore, in our method, we opt to select only one Prefix for each row.

\textbf{Pruning Rules.}
Because of the spatial and temporal relationship definition, the ProSparsity graph is a strict partial ordering set.
Based on previous analysis, we can make simple pruning rules for the graph \revision{to ensure that each node has only one Prefix}.
\begin{itemize}
    \item \revision{First, for each current node, we will traverse all of its Prefix nodes.
    We retain only the edge(s) connected to the Prefix node(s) that have the largest common sub-combination (i.e., the largest subset) with the current node, and remove all other edges.
    That can ensure that the kept Prefix node(s) is the most similar to this current node.}
    \item Then, if the Prefix nodes with the largest subset are multiple, ProSparsity can select any of them. In practice, we keep the \revision{edges from the node with} largest index.
\end{itemize}
After this pruning, the ProSparsity graph will become a directed forest called ProSparsity Forest.

\begin{table}[t]
    \vspace{-10pt}
    \caption{Preliminary experiments on One-prefix and Two-prefix.}
    \vspace{-8pt}
    \centering
    \renewcommand{\arraystretch}{1}
    \resizebox{0.45\textwidth}{!}{
    \begin{tabular}{l|l|c|c}
    \toprule
         & & SpikingBERT SST-2 & VGG-16 CIFAR100 \\ \midrule
        Bit Sparsity &Density & $20.49\%$ & $34.21\%$ \\ \midrule
        \multirow{ 2}{*}{One-Prefix}& Pro Density & $2.98\%$ & $2.79\%$ \\ \cmidrule(lr){2-4}
        &Prefix Ratio & $56\%\times 1$  & $26\%\times 1$ \\ \midrule
        \multirow{ 2}{*}{Two-Prefix} &Pro Density & $2.30\%$ & $1.97\%$ \\ \cmidrule(lr){2-4}
        &Prefix Ratio & $53\%\times 1 + 3\%\times 2$ & $20\%\times 1 + 6\%\times 2$ \\
        \bottomrule
    \end{tabular}
    }
    \label{tbl:prefix}
    \vspace{-10pt}
\end{table}

\textbf{ProSparsity Forest.}
As shown in \Fig{fig:graph_tree}~(c), the ProSparsity Forest will be the fundamental data structure of our design.
After pruning, the ProSparsity Forest may contain multiple directed trees that are independent of each other.
Each tree maintains complete partial ordering information for a processing order constraint of spiking GeMM.
In spiking GeMM, if we process a spike matrix from the top to the bottom, we may not be able to reuse the result of the prefix row since the prefix result is not computed. 
The topology order of the product sparsity tree from root to leaves guarantees that each node comes after its prefix, promising high effectiveness for ProSparsity.

\textbf{Meta Information.}
In practice, ProSparsity Forest is formed by meta information, which is the real data structure in the processing of ProSparsity, \Fig{fig:graph_tree}~(d).
Meta information has two types: spatial information and temporal information, which correspond to the spatial and temporal relationships.
First, the temporal information is a vector with a size of $m$, which records the execution order of each spike row.
The ProSparsity will follow the temporal vector to issue the index of the spike row.
After that, we can find the spatial information according to the issued index.
The spatial information is a list ordered by the row index, and each element in the list contains the Prefix index and ProSparsity pattern.
We can use the spatial information to complete ProSparsity-based computation.
The meta information can ensure the ProSparsity can be processed correctly and efficiently.

Finally, through the \revision{aforementioned} development of the \revision{ProSparsity} data structure and optimization for design space, the time and space complexity of ProSparsity is reduced to an $O(m)$ linear level. 
\section{\name Overview}
\begin{figure}
    \centering
    \includegraphics[width=0.48\textwidth]{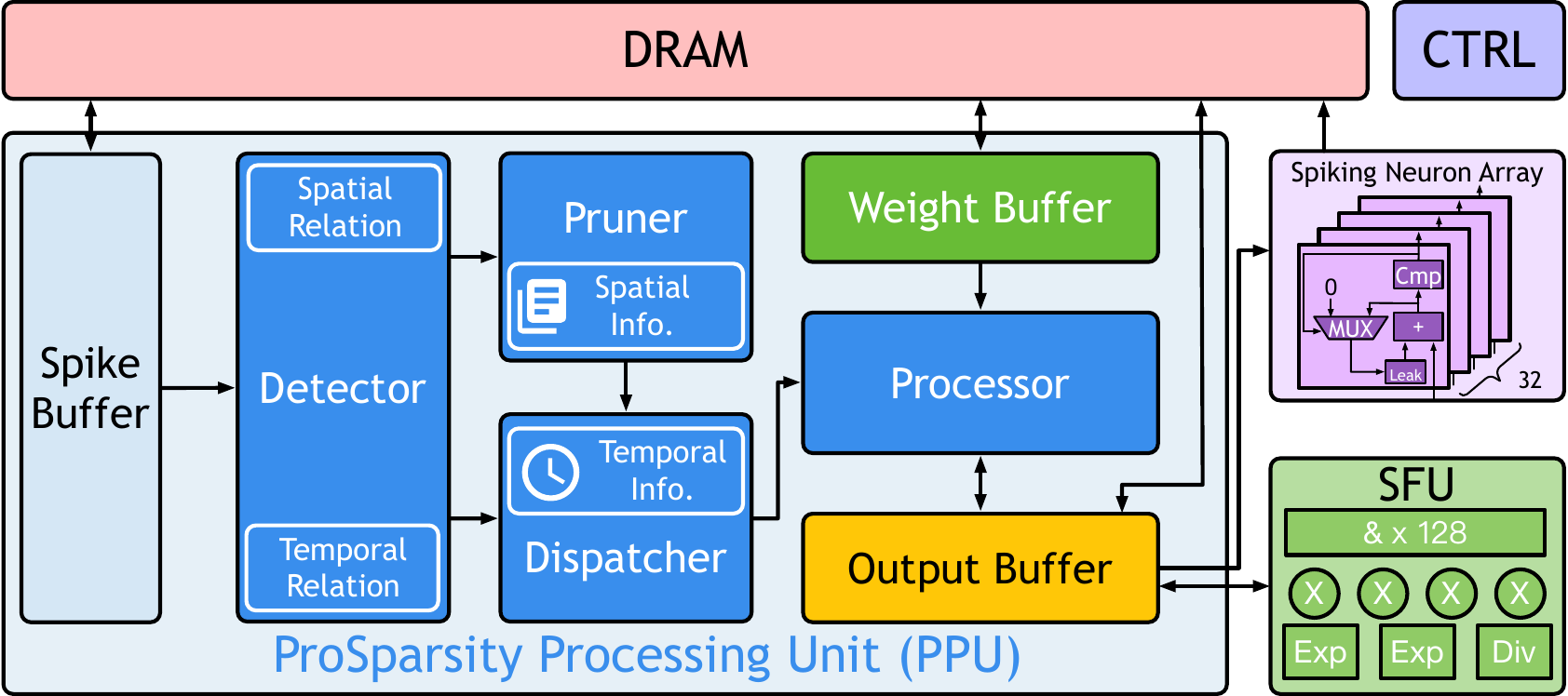}
    \vspace{-8pt}
    \caption{Overview of \name architecture.}
    \label{fig:overview}
    \vspace{-10pt}
\end{figure}

\textbf{Overview.}
\Fig{fig:overview} shows the overall design of \name.
It mainly consists of the ProSparsity Processing Unit (PPU), Spiking Neuron Array, and special function unit (SFU).
\name processes the inference of SNNs layer-by-layer.
The PPU is our core design for processing ProSparsity and matrix computation for each layer.
After that, the results are sent to the Spiking Neuron Array to generate spikes for the next layer, which is a general and necessary unit for SNNs.
In this study, we mainly focus on our proposed PPU design.
Based on the analysis of the prior section, we can divide the whole processing into four stages: relationship detection, spatial information pruning, meta-information generation, and ProSparsity-based computation, which correspond to the \textbf{Detector}, \textbf{Pruner}, \textbf{Dispatcher}, and \textbf{Processor}, respectively.

\textbf{Detector.}
The input of the Detector is the spike matrix stored in the spike buffer.
The Detector can identify the preliminary spatial and temporal relationships for each combination of two spike rows, which can be used to build the ProSparsity Graph.
After that, the Detector will send the original spatial and temporal relationships to the Pruner and Dispatcher to build the ProSparsity Forest.

\textbf{Pruner.}
The Pruner will prune the complex spatial relationship to generate a Prefix for each spike row.
Pruner is also responsible for generating the ProSparsity pattern.
After that, Pruner forms the two pieces of information into meta-spatial information, as shown in \Fig{fig:graph_tree}~(d), and sends them to the Dispatcher.

\textbf{Dispatcher.}
The Dispatcher gathers the spatial information of every spike row and extracts the temporal information according to the temporal relationship.
The Dispatcher can construct the ProSparsity Forest with meta information.
The ProSparsity Dispatcher leverages the meta information to dispatch the instructions and data to the Processor for the computation of ProSparsity-based spiking GeMM.

\textbf{Processor.}
The Processor accesses the weight and output buffer to fetch the corresponding data based on the Prefix and ProSparsity pattern.
Based on the temporal information, the Processor can finish inner product computation in the topology order of ProSparsity Forest and store the result in the output buffer.
When the latter spike row needs the Prefix, the Processor can immediately access it from the output buffer.
After finishing all computations of a spiking matrix, PPU will send the result to the SFU or Spiking Neuron Array.

\textbf{Support for Transformers.}
To support future SNN models, we extend our PPU with an SFU. We reuse the PPU for spiking-GeMM-like operations in spiking attention~\cite{bal2024spikingbert,zhou2022spikformer}. With the assistance of the SFU for exponentiation and multiplication in softmax or layer normalization, we support various spiking transformers~\cite{zhou2022spikformer,bal2024spikingbert,lv2023spikebert,yao2024sdt}.

\section{ProSparsity Processing Unit}
In this section, we design different parts of the ProSparsity Processing Unit based on the insight of \Sec{sec:product sparsity}.

\begin{figure*}
    \centering
    \includegraphics[width=0.9\textwidth]{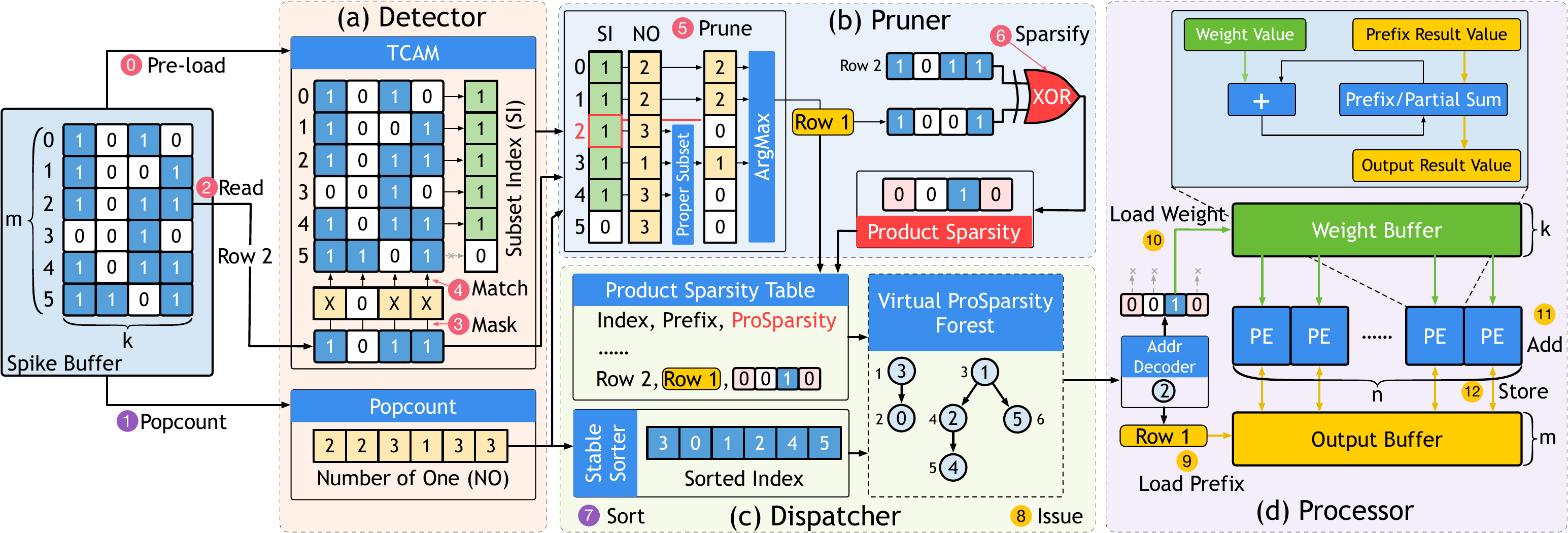}
    \vspace{-8pt}
    \caption{ProSparsity Processing Unit (PPU) is processing the Row~2 in the ProSparsity spiking GeMM with the tiling size of $m\times n\times k$.}
    \label{fig:PU}
    \vspace{-10pt}
\end{figure*}

\subsection{Spiking GeMM Tiling}
\name conducts the tiling method for processing each spiking GeMM in a layer. 
Tiling of matrix multiplication is a commonly used technique in DNN accelerators~\cite{chen2016eyeriss, chen2014diannao, gao2019tangram} that decomposes a large GeMM into several smaller tiles. 
This approach allows the on-chip buffer to efficiently capture the data reuse and reduce the expensive global memory traffic during the computation of large GeMMs.
As shown in \Fig{fig:PU}, we adopt a tiling size of $m \times n \times k$ in \name, where the spike sub-matrix has $m$ rows and $k$ columns, the weight sub-matrix has $k$ rows and $n$ columns, and the output matrix has $m$ rows and $n$ columns. 
Therefore, we set the corresponding buffer size for on-chip buffers and consider a double buffer mechanism to overlap memory access from DRAM with computation.

However, the tiling strategy for ProSparsity is much more important than that in previous DNN/SNN accelerators because the selection of tile sizes can significantly impact its final sparsity and performance.
For example, if we only set $m=1$, apparently, ProSparsity will be invalid because there is no opportunity to reuse a Prefix.
Similarly, the selection of $k$ can also influence the sparsity.
However, we cannot select a very large $m$ or $k$, which will cause excessive area and energy overhead and performance drops.
Fortunately, the number of $n$ has no impact on ProSparsity.

Therefore, we build an end-to-end evaluation model to quantitatively discuss the influence of the tile size in detail in \Sec{sec:influence of tiling}, considering not only the sparsity but also the hardware performance (speedup) based on our complete \name design.
Here, we first present the PPU design in the following sub-sections.

\subsection{ProSparsity Detector}
\textbf{Spatial Detection.}
As discussed in \Sec{sec:eff_dse}, the detection of the ProSparsity is to find the PM and EM relationship for spike rows.
Therefore, the ProSparsity graph requires $O(m^2)$ complexity to finish detection, which introduces too much runtime overhead.
To tackle this challenge, we employ the parallel search ability of Content Addressable Memory (CAM)~\cite{pagiamtzis2006content}, which is a specialized type of memory that searches its entire stored contents in a single clock cycle to find a match to input query data. CAM is commonly used in network routers~\cite{maurya2010dynamic}, data mining~\cite{mining2006introduction}, and various data-intensive applications~\cite{fu2023clap}.
Specifically, we use Ternary CAM (TCAM)~\cite{liu2002routing} to reduce the time complexity to $O(1)$ for each row and $O(m)$ in total for a spike matrix, as shown in \Fig{fig:PU}~(a).
TCAM can match bits with three states: 0, 1, and ``don't care'' ($X$), providing flexibility for fuzzy matching.
Therefore, {we can use the TCAM unit to find all subset indices of a spike row within a single clock cycle.}

Step~\circlednumberpink{0}, when a spike sub-matrix (tile) is sent to PPU, Detector will pre-load a $m\times k$ spike tile into the TCAM with $m$ entries, each entry with $k$ bits.
After initialization, Step~\circlednumberpink{2} and \circlednumberpink{3} will read each spike row (e.g., Row~2) and mask the one value of the row as '$X$' (mask$(1011) = X0XX$).
After that, Step~\circlednumberpink{4}, TCAM will return all matched entries' indices (i.e., Subset Index, SI).
Finally, we realize efficient and parallel spatial information detection with only one cycle.

\textbf{Temporal Detection.}
We use the number of spikes in each row as preliminary temporal information, as shown in \Fig{fig:PU}~(a)~~\circlednumberpurple{1}.
To efficiently obtain the number of spikes, we introduce the Popcount~\cite{intel2007sse4} operation, which counts the number of ones (NO) in a binary sequence.
We place multiple lightweight popcount units in the Detector to obtain this preliminary information.

\subsection{ProSparsity Pruner}
The Pruner prunes the spatial relationships to a single Prefix for each row according to our pruning rules.

\textbf{Efficient Pruning.}
The Detector provides spatial information (Subset Index, SI) with only subset information without considering the partial ordering underlying the temporal relationship (Number of One, NO).
As we discussed in \Sec{subsec:tr}, for two rows with Exact Match (EM), only the row with the smaller index should be as the Prefix.
Therefore, as shown in \Fig{fig:PU}~(b)~\circlednumberpink{5}, we design a proper subset filter for the spike rows with a larger index than the existing query row to eliminate violations according to the partial ordering.
Combining SI and NO vectors, we can directly eliminate the EM row with a larger index.
For instance, we remove Row~4 from the Prefix candidates for Row~2, where Row~4 has a larger index and is EM to Row~2.
After that, we employ an Argmax unit to select only one Prefix (Row~1) for Row~2, based on the heuristic pruning rules in \Sec{sec:eff_dse}.

\textbf{ProSparsity Sparsifying.}
Consequently, we can utilize the selected Prefix row to generate the ProSparsity pattern, which contains the spike values to be calculated.
We use only one XOR (exclusive or) unit to get the ProSparsity for the query row by conducting a bit-wise XOR operation in Step \circlednumberpink{6}.
For example, the Row~2 with the Prefix Row~1 can get the ProSparsity pattern: $1011\oplus 1001 = 0010$.
Because the Prefix row is a subset of the query row, the bit-wise XOR is equivalent to the operation $S_q - S_p$.

The spatial information for a query row is now completed and will be sent to the Dispatcher for subsequent usage.

\subsection{ProSparsity Dispatcher} \label{sec:dispatcher}
The ProSparsity Dispatcher serves as a link between each part of the unit in PPU, as shown in \Fig{fig:PU}~(c).
It gathers spatial information from the Pruner and stores it in the product sparsity table. 
It also extracts temporal information according to temporal relationships from the Detector. 
The Dispatcher extracts the execution order and then dispatches tasks to the Processor.
However, there are still some difficulties in generating execution orders under the space and time trade-offs.

\textbf{Memory Space Issue.}
Generally, the spatial information contains the complete partial ordering for the ProSparsity Forest.
If we record the Prefix and Suffix for each row, we can traverse the Forest to generate an execution order using Breadth or Depth First Search.
However, because one spike row may have multiple Suffixes, we should use a matrix with the shape of $m\times m$ to record the two-row spatial information of $m$ rows.
Furthermore, the Prefix-Suffix matrix has a large density, as shown in \Tbl{tbl:prefix}.
Even if we adopt a sparse format to store the matrix, the space complexity is still $O(m^2)$ and will lead to irregular accesses. 
Therefore, the $O(m^2)$ memory space will significantly enlarge the on-chip area more than 10 times compared to the existing design, which is unacceptable for our design.
Fortunately, there is only one Prefix for each row.
We exploit this feature of ProSparsity Forest to store the spatial information in the product sparsity table with $O(m)$, as shown in  \Fig{fig:PU}~(c).
Therefore, we eliminate the suffix information in the tree to achieve space efficiency.

\textbf{Search Time Issue.}
However, that will cause the search time issue when we lack Suffix information, i.e., we cannot run Breadth or Depth First Search in a $O(m)$ time.
We need to traverse the product sparsity table and then traverse the Forest from the leaf to the root node with $O(m\times d)$ time, where $d$ is the depth of the Forest.
In \Sec{sec:ablation}, our ablation study shows this way will slow down \name by 32\%.

\textbf{Efficient Generation.}
We propose an interesting and efficient solution to obtain the execution order faster without searching in the product sparsity table. This solution is based on two rules of our design:
\begin{itemize}
    \item For the Partial Match, the Prefix is the proper subset of the Suffix, i.e., the number of one (NO) in the Prefix is smaller than the Suffix.
    \item For the Exact Match, the Prefix and Suffix have the same NO. The Prefix has a smaller index than the Suffix.
\end{itemize}
Therefore, we can use a stable sorting approach to generate the temporal information, as shown in \Fig{fig:PU}~(c)~\circlednumberpurple{7}.
After sorting, we can ensure that the Prefix comes before the Suffix with the temporal information, which is easy to prove.
We implement the parallel stable sorter, using a bitonic sorter~\cite{batcher1968sorting} with $O(log_2^2 m)$ time complexity and $O(m)$ space complexity, which can concurrently conduct the sorting with the Detector and Pruner.
In this way, the spatial and temporal information can be completed in the Dispatcher and form a virtual ProSparsity Tree, as shown in \Fig{fig:PU}~(c).

Then, the Dispatcher issues tasks (\circlednumberorange{8}) to the Processor according to the meta information.

\subsection{Product Sparsity Processor}
The Processor conducts ProSparsity-based matrix computation according to the guidance of the Dispatcher.

\textbf{ProSparsity Row-wise Dataflow.}
\name employs a row-wise dataflow, i.e., a spike row with the size of $k$ is processed at a time, and a corresponding output row with the size of $n$ is generated, involving $n\times k$ weight.
The Processor processes spike rows in the order provided by the temporal information in the Dispatcher.
This order guarantees that the Prefix row in the output matrix is already generated when processing the Suffix row.
For each spike row, the Prefix row in the output matrix is fetched and serves as a starting point of the partial sum.
Then, we examine the position of the spikes. 
For each spike, we fetch the corresponding row in the weight matrix and accumulate it on the partial sum.
The summed result is written to the corresponding row in the output matrix.

\textbf{Efficient dataflow implementation.}
Our Processor design realizes the aforementioned dataflow, as shown in \Fig{fig:PU}~(d).
It mainly consists of a PE array for vectorized accumulation and an address decoder to get data addresses for weight and output buffer.
In the process of each row, the Prefix row result is loaded to the partial sum register in the PE array, as shown in Step \circlednumberorange{9} in \Fig{fig:PU}.
Step \circlednumberorangetiny{10} is performed and controlled by the address decoder.
This weight address is decoded by bit scan forward~\cite{intel2023manual} operation, which finds the index of the first 1 in the ProSparsity pattern. 
The decoder then flips this found bit to 0.
Step \circlednumberorangetiny{11} is performed by the PE array for accumulation.
Step \circlednumberorangetiny{10} and Step \circlednumberorangetiny{11} are repeated until all the 1s in the spike row are consumed, and the result is written back (\circlednumberorangetiny{12}) to the output buffer.
Since the final result of a tile in the output matrix is the summation of multiple results, this result will also accumulate on the partial sum of the output tile.
Our design efficiently bypasses all the zeros in the ProSparsity pattern, supporting unstructured sparsity.

\section{Pipeline Scheduling}
We design a pipeline scheduling mechanism to maximize the utilization of our hardware in PPU. 
A mini example of a pipeline with 4 rows is shown in \Fig{fig:pipeline}.
The whole PPU can be divided into two phases: the ProSparsity processing phase and the computation phase.
The Detector, Pruner, and Dispatcher are responsible for ProSparsity processing, while the Processor conducts the computation phase. 
These two phases are conducted in a tile-by-tile and sequential manner.

\subsection{Intra-phase Pipeline}
\begin{figure}
    \vspace{-10pt}
    \centering
    \includegraphics[width=0.48\textwidth]{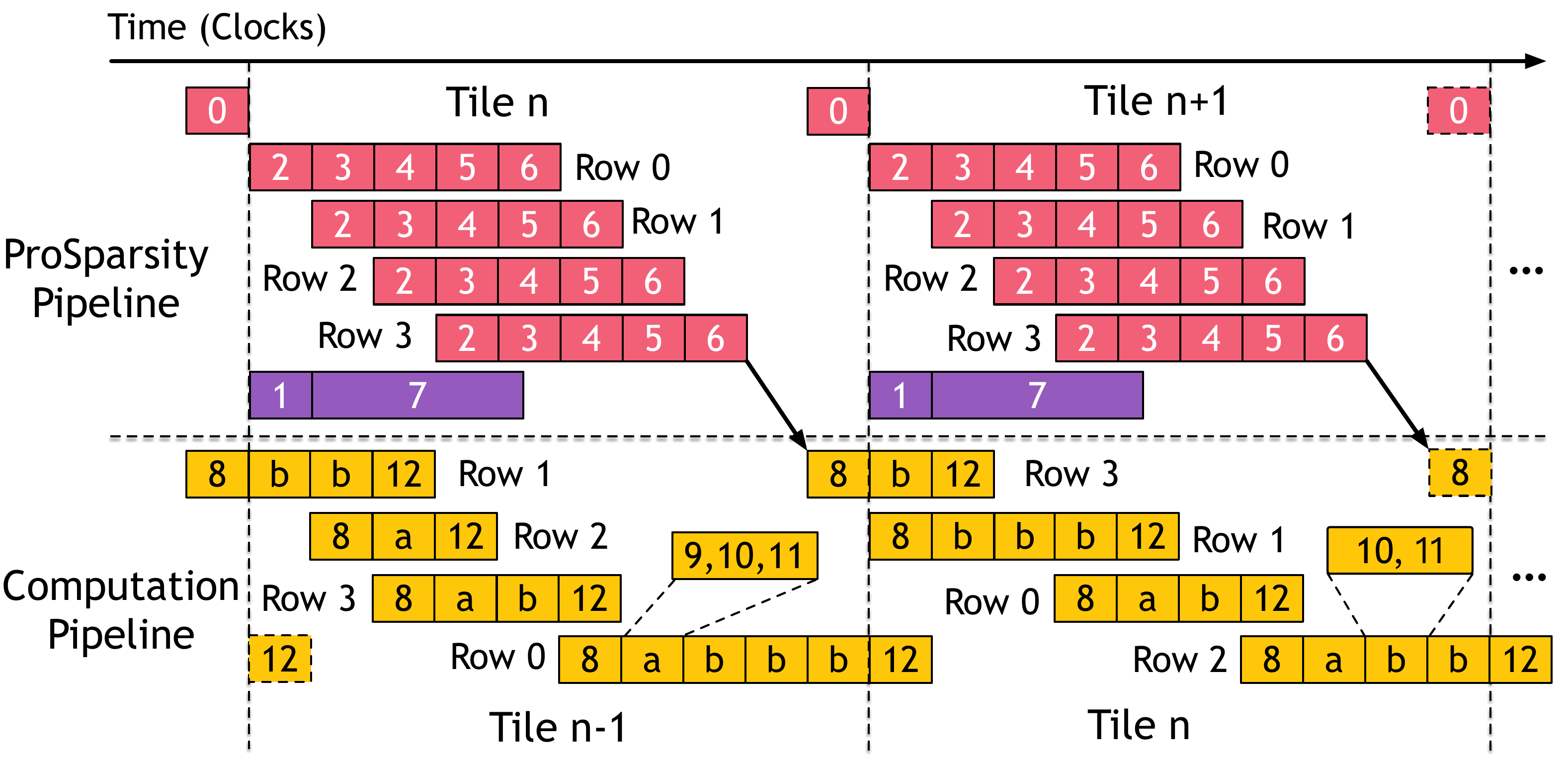}
    \vspace{-8pt}
    \caption{ProSparsity pipeline scheduling.}
    \label{fig:pipeline}
    \vspace{-15pt}
\end{figure}

The Step \circlednumberpink{2} to \circlednumberpink{6} in Detector, Pruner, and Dispatcher is fully pipelined, as illustrated in \Fig{fig:pipeline}.
Therefore, this pipeline has five stages, achieving the throughput of 1 instruction (spike row) per cycle.
Therefore, when the tile has $m$ rows, we only need $m+4$ cycles for all the spatial information.
The temporal information could be extracted (Step \circlednumberpurple{1} and \circlednumberpurple{7})  concurrently with Step \circlednumberpink{2} to \circlednumberpink{6} pipeline due to our observation that decouples the spatial and temporal information.
The Step \circlednumberpurple{7} takes the parallel bitonic sorter with $O(log_2^2 m)$ time complexity to get the sorted temporal information. 
Thus, the cycles of sorting are far less than $m$.
The pre-load Step \circlednumberpink{0} can perform concurrently through a double-buffer-based TCAM design.
Therefore, the overall cycle for a ProSparsity processing phase is $m+4$.

For the computation phase, our target is to maximize the utilization of the PE array.
The Processor also has four stages in the pipeline: Step \circlednumberorange{8}: issue, Step \circlednumberorange{9} and \circlednumberorangetiny{10} for decode and load, Step \circlednumberorangetiny{11} is execute, and Step \circlednumberorangetiny{12}: write back.
For simplicity, we present them into `a' and `b', as depicted in \Fig{fig:pipeline}.
For each spike row, the computation phase may have different cycles due to the spike row having different sparsity.
Therefore, for $m$ rows, the overall cycle for a computation processing phase is $\ge m+4$.

\subsection{Inter-phase Pipeline. }
We design an inter-phase pipeline to overlap these two phases.
The ProSparsity processing phase is in Tile $n$, while the Processor runs on Tile $n-1$ with the meta information already gathered in the Dispatcher. 
Therefore, the product sparsity table for spatial information adopts a double-buffer design to enable the two-phase overlapping.
Based on intra-phase pipeline designs,  the cycle of the ProSparsity processing phase is always less than the computing phase.
Thus, the ProSparsity processing phase of a tile is perfectly overlapped by the computation phase of the previous tile.
Since we conduct tiling and split a large spiking GeMM into multiple tiles, the computation phase can cover almost all of the ProSparsity processing phases except for the first tile phase, which has a minor impact.

Finally, \name can realize an overhead-free ProSparsity processing and fully utilize the computation units with the pipeline scheduling.

\section{Evaluation}

\subsection{Methodology}

\textbf{SNN models and dataset.} We evaluate \name on spiking CNNs that previous SNN ASICs design targets.
We also evaluate \name on different spiking transformers that prior SNN accelerators do not efficiently support.
For model architecture, we select VGG~\cite{simonyan2014very} and ResNet~\cite{he2016deep} for spiking CNNs and SpikeBERT~\cite{lv2023spikebert}, SpikingBERT~\cite{bal2024spikingbert}, Spikformer~\cite{zhou2022spikformer}, SDT~\cite{yao2024sdt} for spiking transformers.
We use the default configuration for a number of layers, dimensions, and time steps in the paper.
These SNN models run on NLP or CV datasets, including SST-2, SST-5, MR, QQP, MNLI, CIFAR10, CIFAR100, CIFAR10-DVS~\cite{socher2013recursive,pang2005seeing,wang2018glue,williams2017broad,krizhevsky2009learning,li2017cifar10}.
To get the spike activation pattern in each layer, we run these models implemented in PyTorch.
We extract the runtime information and use it in our experiment.
\revision{\name yield iso-accuracy compared to the model in PyTorch as our proposed ProSparsity is an algorithm-agnostic and lossless method.}

\textbf{Hardware Configuration.}
\name configuration is shown in \Tbl{tab:config}. We implement our core architecture in SystemVerilog~\cite{ieee2017systemverilog}, including PPU, spiking neuron array, and SFU. The bitwidth of weights is set to 8 bits, which aligns with previous ASICs and neuromorphic chips. We use Synopsys Design Compiler to synthesize our RTL logic to get area and power with ARM's standard cell library under a commercial 28nm process. The spike, weight, and output buffer are evaluated by CACTI 7.0~\cite{balasubramonian2017cacti} under a 28nm process to get area and power. The off-chip DRAM power is simulated by DRAMsim3~\cite{li2020dramsim3}.

\begin{table}[b]
    \vspace{-10pt}
    \caption{\name Architecture Setup.}
    \vspace{-8pt}
        \centering
        \begin{tabular}{l|l}
            \toprule
            Tile Size & $m=256$; $n=128$; $k=16$.\\ \midrule
            Detector & 1KB TCAM; 8 Popcounts. \\ \midrule
            Pruner & 256 Channel Subset Unit \& Argmax. \\ \midrule
            Dispatcher & 1.5KB Product Sparsity Table. \\ \midrule
            Processor & 128 PEs 8-bit Add, $n=128$.\\ \midrule
            Spiking Neuron Array & 32 Cells LIF Neuron.\\ \midrule
            Special Function Unit & 128AND/OR; 32MUL; 8EXP; 1DIV. \\ \midrule
            On-chip Buffer size & 8KB Spike; 32KB Wgt; 96KB Out. \\ \midrule
            DDR4 DRAM & 4Gb $\times$ 16 2133R 4 Channels 64GB/s. \\
            \bottomrule
        \end{tabular}
        \label{tab:config}
\end{table}

\textbf{Architecture Modeling and Baselines.}
We build a cycle-accurate simulator to emulate the behavior of \name and evaluate the runtime performance.
\name is benchmarked against dense DNN accelerator Eyeriss~\cite{chen2016eyeriss}, SNN accelerators PTB~\cite{lee2022parallel}, SATO~\cite{liu2022sato}, \revision{and MINT~\cite{yin2024mint,yin2022sata}}, algorithm-modified SNN accelerator Stellar~\cite{mao2024stellar}, and NVIDIA A100 GPU.
We compare their configuration and performance on VGG-16 with CIFAR100 in \Tbl{tab:baselines}. We maintain the same frequency in every accelerator, while Eyeriss and Stellar have 31\% more PEs than SATO, PTB, \revision{MINT}, and \name in their design. 
We simulate Eyeriss, SATO, PTB, and \revision{MINT} under our simulator framework. We also simulate the 
For the performance of Stellar~\cite{mao2024stellar}, we use the statistics reported in their paper since its algorithm modification is not open-sourced, and we are not able to access its sparsity pattern.
Since previous SNN accelerators only support the linear layer (e.g., QKV projection, output projection, feed-forward layer) in spiking transformers but cannot deal with attention and layer normalization, we run PTB, SATO, and \revision{MINT} on the linear layer of spiking transformers for a comparison.
We further compare with NVIDIA A100 GPU~\cite{nvidia_a100} with 80~GB global memory, 
\revision{running SNN models using PyTorch and SpikingJelly~\cite{spikingjelly}, measuring the end-to-end inference time and power by averaging over the whole dataset.}
\revision{Additionally, we include the state-of-the-art SNN accelerator LoAS~\cite{yin2024loas} in our analysis, which focuses on weight pruning and induces dual-side sparse computation. 
We implement LoAS's algorithm to provide a comprehensive sparsity analysis comparing LoAS and \name.}

\begin{table}
    \vspace{-10pt}
    \centering
    \caption{
    Compare \name with prior accelerators on VGG-16.}
    \vspace{-8pt}
    \resizebox*{0.5\textwidth}{!}{
    \begin{tabular}{l|c|c|c|c|c|c}
    \toprule
    ~ & Eyeriss~\cite{chen2016eyeriss} & SATO~\cite{liu2022sato} & PTB~\cite{lee2022parallel} & \revision{MINT~\cite{yin2024mint}} & Stellar~\cite{mao2024stellar} & \name \\ \midrule
    Frequency & 500MHz & 500MHz & 500MHz & \revision{500MHz} & 500 MHz& 500MHz \\\midrule
    Technology & 28nm & 28nm & 28nm & \revision{28nm} & 28nm &  28nm \\ \midrule
    PEs & 168 & 128 & 128 & \revision{128} & 168 & 128 \\ \midrule
    ALU & 8-b MAC & 8-b Add & 8-b Add & \revision{2-b Add} & 12b Add &  8-b Add\\ \midrule
    Area (mm\textsuperscript{2}) & 1.068 & 1.13 & - & \revision{-} & 0.768 & \textbf{0.529} \\ 
    \midrule
    Throughputs & 29.40 & 33.63 & 41.37 & \revision{62.07} & 190.44 & \textbf{390.10} \\
    (GOP/s) & (1.00$\times$) & (1.14$\times$) & (1.41$\times$) & \revision{(2.11$\times$)} & (6.48$\times$) &  (\textbf{13.27$\times$}) \\ \midrule
    Energy Effi. & 16.67 & 49.70 & \revision{34.15} & 75.61 & 142.98 & \textbf{299.80} \\
    (GOP/J) & (1.00$\times$) & (2.98$\times$) & \revision{(2.05$\times$)} & (4.53$\times$) &  (8.57$\times)$ &  (\textbf{17.98$\times$}) \\ \midrule
    Area Effi. & 27.53 & 29.76 & \multirow{2}{*}{-} & \revision{\multirow{2}{*}{-}} & 247.97 &  \textbf{737.17} \\
    (GOP/s/mm\textsuperscript{2}) & (1.00$\times$) & (1.08$\times$) & ~ & ~ &(9.01$\times$) &  (\textbf{26.78$\times$}) \\
    \bottomrule
    \end{tabular}
    }
    \label{tab:baselines}
    \vspace{-10pt}
\end{table}

\subsection{Tiling Exploration} \label{sec:influence of tiling}

\begin{figure}[t]
    \centering
    \includegraphics[width=0.48\textwidth]{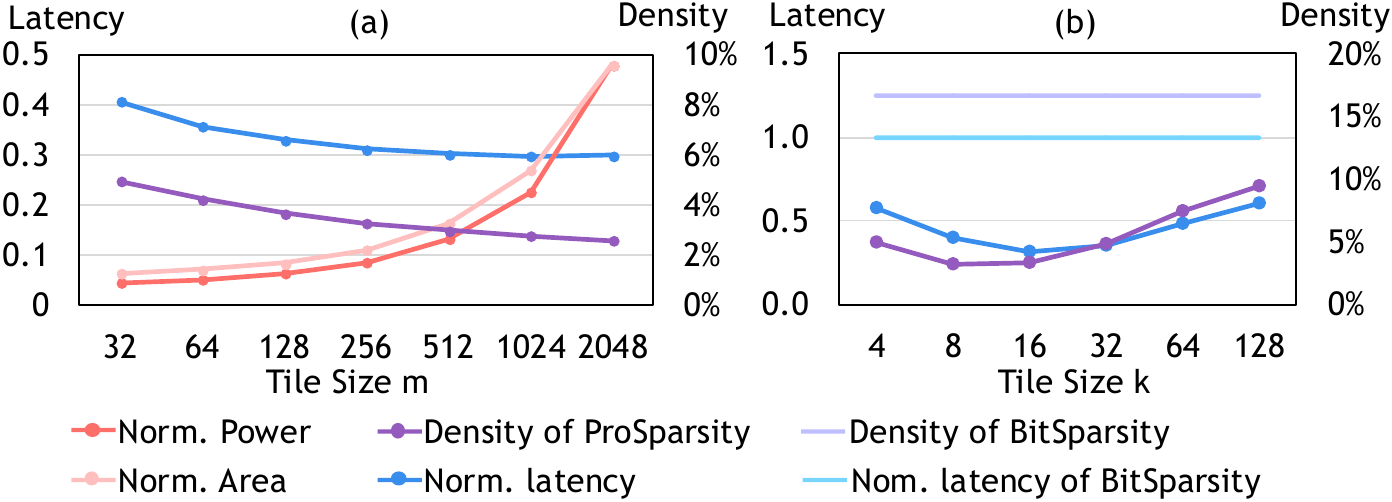}
    \vspace{-8pt}
    \caption{
    \revision{Performance change with tile size $m$ (left) and $k$ (right).}}
    \label{fig:tile_size}
    \vspace{-10pt}
\end{figure}

\begin{figure*}
    \centering
    \vspace{-8pt}
    \includegraphics[width=0.98\textwidth]{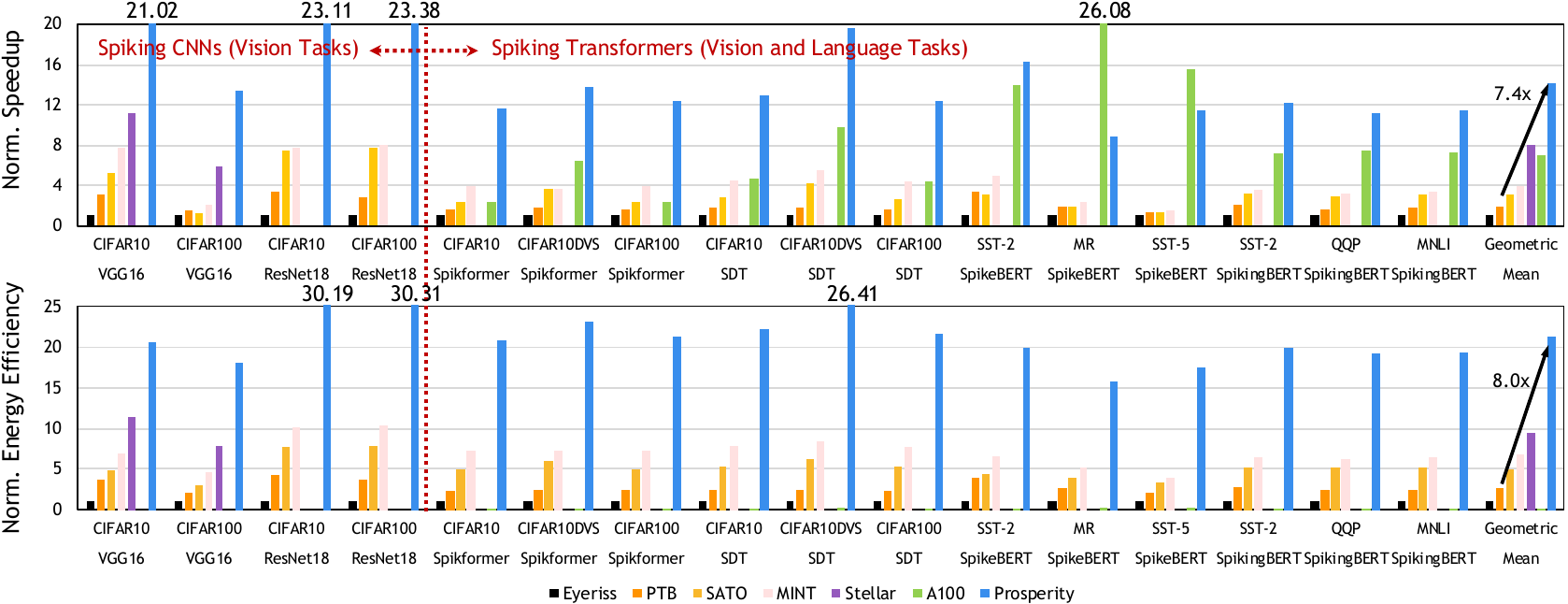}
    \vspace{-10pt}
    \caption{
    \revision{Speedup and energy efficiency (normalized by Eyeriss) of \name. \name support both spiking CNNs and spiking transformers.}}
    \label{fig:perf}
    \vspace{-10pt}
\end{figure*}
We observe that the appearance of product sparsity is related to the tile size of the spike matrix.
Therefore, we conduct design space exploration on tile sizes to maximize product sparsity while maintaining a reasonable on-chip buffer size.
\Fig{fig:tile_size} shows the relationship between latency/density and the spike tile size in \name.
The latency is normalized to the latency of the bit sparsity baseline.
The statistics are averaged across all spiking CNNs and spiking transformers.
We observe that larger $m$ always leads to higher product sparsity, while a suitable $k$ leads to the highest product sparsity.
\revision{In considering the optimal tile size $m$, we must balance latency improvements against hardware costs. While a larger $m$ generally reduces latency, it also increases hardware overhead for on-chip buffers and TCAM. The pink/coral lines in \Fig{fig:tile_size} illustrates the relationship between area/power and tile size in \name. As $m$ increases, hardware overhead grows super-linearly. Therefore, we should carefully weigh the diminishing returns on latency against the escalating hardware costs when selecting the tile size.} 

On the $M$ dimension, a larger $m$ means more rows within a tile, providing a larger scope for searching rows that can be used to exploit product sparsity.
For each row in the spike matrix, they can look up more rows for the prefix, maximizing the product sparsity.

On the $K$ dimension, a larger $k$ means more columns within a tile, making the spike set of each row bigger and more complicated and, therefore, harder to find a Prefix.
When each spike set is small and simple, rows that satisfy equal or subset relationships are more likely to be found.
However, as the tile size $k$ gets extremely small, e.g., 4, more rows will appear that have less than two spikes.
Exploiting product sparsity for rows that have zero or one spike is meaningless as it does not need accumulation.

According to this exploration, we select a proper tile size for the spike matrix: $m = 256$ and $k = 16$.

\subsection{End-to-End Performance Analysis}
\Fig{fig:perf} shows the speedup and energy efficiency of \name compared with baselines running different models and datasets.
We can also observe superior throughputs and area efficiency of \name from the statistics in \Tbl{tab:baselines}. We will compare with other baselines and elaborate on the reasons for our superiority.

\textbf{Comparison with ASICs.}
Eyeriss serves as a baseline accelerator that processes SNNs in a dense manner.
\name achieves a $14.2 \times$ speedup and a $21.4 \times$ energy efficiency compared with Eyeriss.

\name achieves $7.4 \times$ and $4.8 \times$ speedup, $8.0 \times$ and $4.2 \times$ energy efficiency over PTB and SATO for the following two main reasons.
These two accelerators trade the sparsity for parallelism when processing spiking GeMM.
PTB forces the spike matrix pattern into a structured sparsity during spiking GeMM computation. It processes every bit within a time window as long as one spike exists in the window.
This indicates that some zeros are not skipped in PTB.
SATO distributes spike rows to multiple PE groups, thus suffering from workload imbalance problems.
\name has conquered these two problems through our dedicated Processor design.
\name's row-wise dataflow with address decoder identifies the spikes and skips every zero, efficiently processing unstructured spike patterns in a single PE array.
Significantly, \name further reuses the prefix result to skip a significant amount of computation through ProSparsity.
\revision{\name achieves a $3.6 \times$ speedup and $3.1 \times$ energy efficiency over MINT. While MINT leverages BitSparsity and quantization (weights and membrane potentials), ProSparsity approach provides complementary benefits, enabling superior performance.}

Compared with the algorithm-hardware co-design accelerator Stellar, \name also achieves a $2.1 \times$ speedup and $2.2 \times$ energy efficiency on spiking CNNs even when Stellar has 31\% more PEs.
The speedup mainly comes from our significant improvement in sparsity.
Stellar improved sparsity by modifying the SNN algorithm,
while our proposed product sparsity improves the sparsity to a greater extent and in an algorithm-agnostic manner.

\textbf{Comparison with A100.}
We compared with A100 on the end-to-end performance of various spiking transformers.
Because of the distinctive and diverse operations of spiking transformers,
spiking transformers are not efficiently supported by previous ASICs~\cite{lee2022parallel,mao2024stellar} and neuromorphic chips~\cite{davies2018loihi}.
Therefore, GPU is currently the better hardware for spiking transformers.
\name can achieve a $1.79 \times$ speedup even though A100 has 312 TOPS peak throughput and 826 $mm^2$ area that is far more than \name with only 0.529 $mm^2$.
Notice that \name has no significant speedup over A100 on SpikeBERT.
This is because SpikeBERT is a relatively large model with 12 blocks of transformer encoder and 768 hidden dimension size.
The minor speedup is caused by the better utilization of the computation core on A100.
Nevertheless, \name stands out in the competition for energy efficiency. \name has $193 \times$ energy efficiency improvement over A100.
The reasons for this huge lead are twofold.
A100's tensor core is designed for GeMM in DNN, which performs highly parallel MACs.
However, spiking GeMM in SNN requires only accumulation, resulting in the underutilization of computation units in the tensor core.
Moreover, \name efficiently skips the zeros in spiking GeMM through bit sparsity and product sparsity, while GPU's SIMT architecture is rigid and unable to eliminate these redundancies.

\begin{figure}
    \centering
    \vspace{-8pt}
    \includegraphics[width=0.48\textwidth]{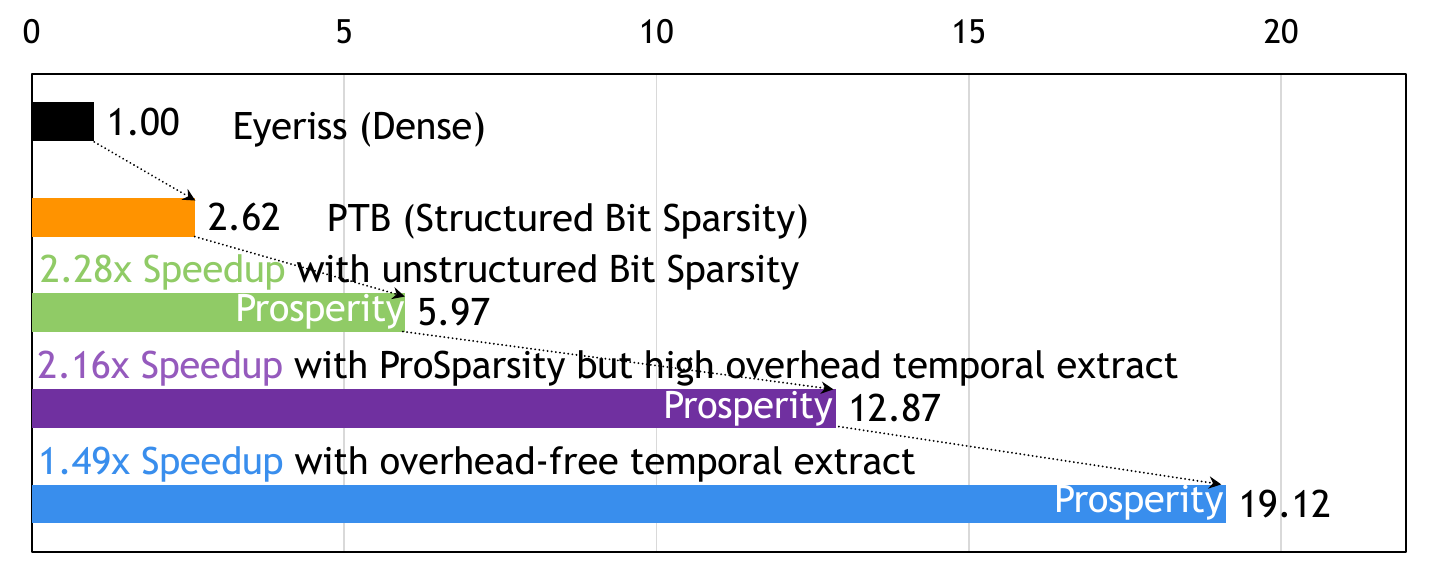}
    \vspace{-10pt}
    \caption{Ablation study of \name.}
    \label{fig:ablation}
    \vspace{-10pt}
\end{figure}

\subsection{\revision{Ablation Study: Speedup Analysis}} \label{sec:ablation}
We conduct an ablation study of each part of our design.
This study is conducted on all evaluated models with average results.
\name efficiently handles the unstructured bit sparsity in spiking GeMM through our row-wise dataflow and an efficient address decoder.
This unstructured sparsity processing achieves a $2.28\times$ speedup compared to PTB, which conducts structured sparsity processing.

We then introduce product sparsity in the accelerator but use the high-overhead Dispatcher method explained in \Sec{sec:dispatcher}.
This implementation utilizes product sparsity to reduce the number of accumulations but induces extra overhead to process rows in the correct order.
It further achieves a $2.16\times$ speedup compared to simply using bit sparsity.
However, this implementation does not fully unleash the potential of product sparsity.

Our proposed overhead-free dispatch elegantly removes the overhead of finding the processing order through our keen observation.
It further achieves a $1.49\times$ speedup compared to the high overhead dispatching.
Finally, compared to bit sparsity, \name achieves an average of $3.2\times$  speedup for all models and up to $7.4\times$ speedup on SpikeBERT on SST-5.

\begin{figure}
    \centering
    \includegraphics[width=0.48\textwidth]{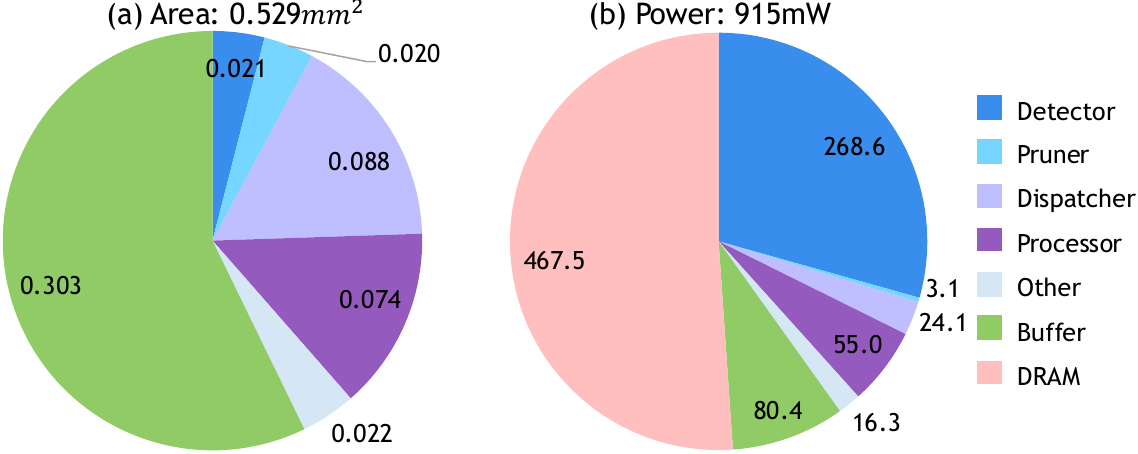}
    \vspace{-8pt}
    \caption{\name area and power breakdown.}
    \label{fig:energy_breakdown}
    \vspace{-12pt}
    \end{figure}

\subsection{\revision{Area and Power Analysis}}
\textbf{Area.}
The area and power overhead of different parts of \name is shown in \Fig{fig:energy_breakdown}.
For the area, we notice that the Dispatcher is the dominant part in \name, except for the on-chip buffer. This is mainly attributed to the product sparsity table that maintains the product sparsity information for each row.
The Processor with PE array also occupies the second largest part of the area.

\textbf{Power.}
The power is evaluated on Spikformer with CIFAR10.
DRAM power takes a large proportion of the total power.
For on-chip power, we observe that the proportion of each part differs from the proportion of the area.
The Detector with TCAM consumes most of the on-chip power because every cell in TCAM is activated to perform a highly parallel search every cycle, while the product sparsity table with a large area is only partially activated each cycle.
The on-chip buffer and Processor also consume considerable energy.
PPU (Detector, Pruner, Dispatcher, Processor) consumes most of the energy of \name's computation unit.
This is reasonable since spiking GeMM is the bottleneck of SNNs, and we allocate most of the resources to spiking GeMM.

\subsection{\revision{Sparsity Analysis}} \label{sec:sparsity_comparison}

\begin{figure}[t]
    \centering
    \includegraphics[width=0.48\textwidth]{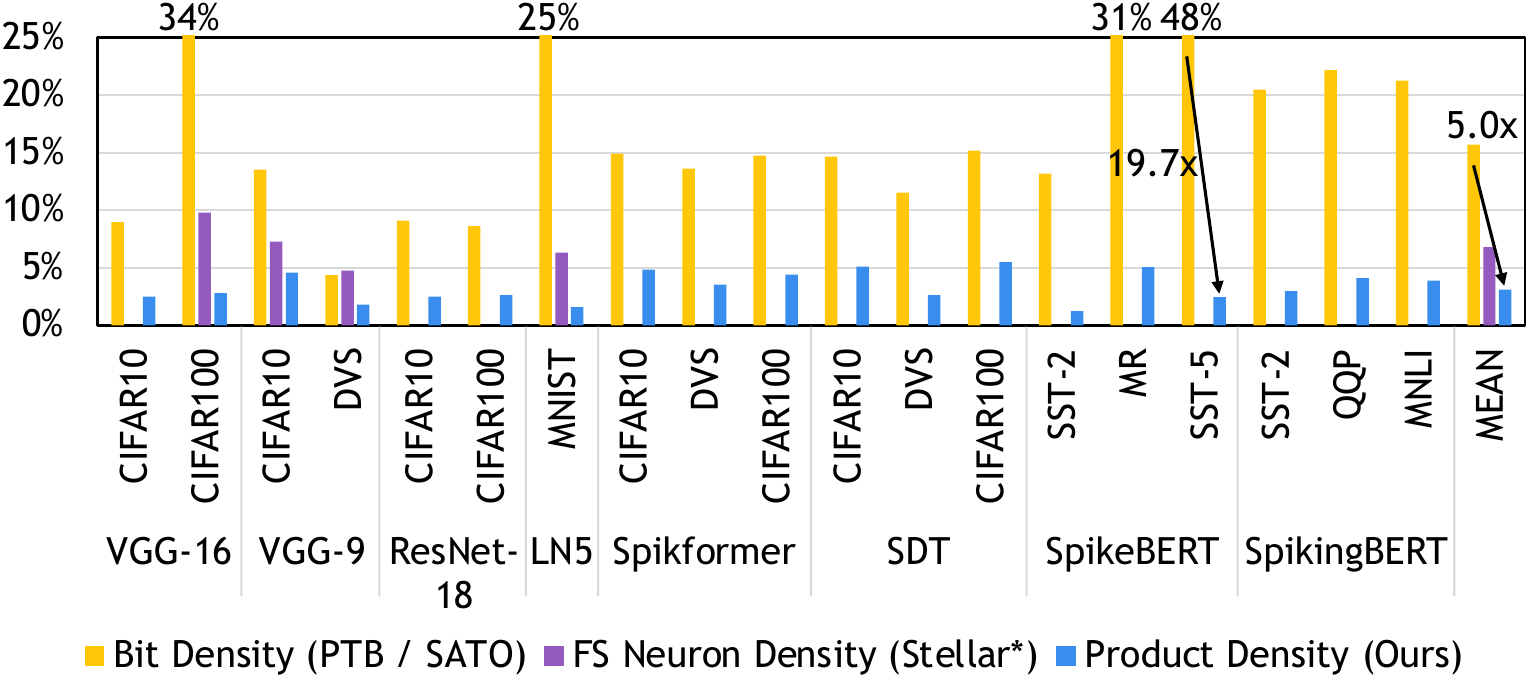}
    \vspace{-8pt}
    \caption{Density comparison of different methods.}
    \label{fig:density}
    \vspace{-12pt}
\end{figure}

\textbf{\revision{Activation-only Sparsity.}}
\Fig{fig:density} compares the density in bit sparsity, Stellar's FS neuron sparsity, and our product sparsity. 
Stellar introduced the use of FS neurons in SNNs to enhance sparsity, but this method sacrifices some accuracy of SNNs and has not been evaluated for applicability to emerging spiking transformers.
In contrast, our product sparsity is algorithm-agnostic and can be applied to any SNN.
Our product sparsity achieves up to a $19.7\times$ and an average $5.0\times$ reduction in density compared to bit sparsity. 
Therefore, ProSparsity on \name brings a $3.2 \times$ speedup compared to BitSparsity-only on \name. 
There is a gap between the theoretic limit of $5.0 \times$ speedup since EM ProSparsity has 100\% sparsity but still takes one cycle to process.
It also achieves an average $3.2\times$ density reduction compared to the algorithmic FS neuron method.
Our product sparsity is effective for spike matrices with arbitrary density. Whether for relatively dense workloads (e.g., SpikeBERT) or relatively sparse workloads (ResNet18), we are able to reduce the density below 5\%.

\textbf{\revision{Dual-side Sparsity.}}
\revision{We demonstrate the effectiveness of ProSparsity when applied to dual-side sparse SNN models proposed in LoAS. 
While LoAS focuses on BitSparsity and pruned sparse weights in SNNs, our ProSparsity introduces a novel, orthogonal direction for optimization. This complementary nature allows ProSparsity to be applied to LoAS-based accelerators, further reducing computation in dual-side sparse SNNs.
To evaluate this synergy, we tested ProSparsity on three spiking CNNs pruned by LoAS. 
We measured three key metrics: weight density, original activation density, and activation density after applying ProSparsity. 
The results are presented in \Tbl{tab:loas}.
LoAS effectively pruned the weights of these models to a density below 5\% with minimum accuracy loss.
By applying ProSparsity to these pruned SNNs, we achieved a significant reduction in activation density, reducing the activation density by $4.1 \times$ on average.}

\begin{table}[b]
    \centering
    \vspace{-15pt}
    \caption{
    \revision{Density of weight and activation in LoAS~\cite{yin2024loas} with ProSparsity.}}
    \vspace{-8pt}
    \resizebox*{0.4\textwidth}{!}{
    \begin{tabular}{l|c|c|c|c}
    \toprule
        Model & Tensor & LoAS & LoAS + \name & Ratio \\ \midrule
        \multirow{ 2}{*}{AlexNet} & Weight &  1.8\% & 1.8\% & - \\ \cmidrule(lr){2-5}
         & Activation &  29.32\% & 9.12\% & $3.21 \times$ \\ \midrule
        \multirow{ 2}{*}{VGG-16} & Weight &  1.8\% & 1.8\% & - \\ \cmidrule(lr){2-5}
         & Activation &  31.07\% & 7.68\% & $4.05 \times$ \\ \midrule
        \multirow{ 2}{*}{ ResNet-19} & Weight &  4.0\% & 4.0\% & - \\ \cmidrule(lr){2-5}
         & Activation &  35.68\% & 6.96\% & $5.13 \times$ \\ \midrule
    \end{tabular}
    }   
    \vspace{-10pt}
    \label{tab:loas}
\end{table}

\subsection{
\revision{Cost Trade-off Analysis in \name}}
\revision{
The overhead of SNN sparsity processing is generally necessary for all DNN accelerators.
For instance, Stellar's sparsity processing overhead is 47\% of its total energy and the computation is only 7\%~\cite{mao2024stellar}.
\name architecture employs ProSparsity processing to reduce floating-point computations in spiking GeMM, introducing a trade-off between processing overhead and computational savings. 
Here, we provide a quantitative analysis, demonstrating that the benefits of ProSparsity outweigh its costs.
The ProSparsity processing phase for a single tile involves:
}

\begin{itemize}
    \item \revision{Bitwise operations in TCAM: $m^2 \times k$.}
    \item \revision{Integer comparisons in the stable sorter: $2m \times \log(m)$.}
    \item \revision{Integer comparisons in the pruner: $m + \log(m)$.}
\end{itemize}

\revision{
Consequently, TCAM bitwise operations have $O(m^2)$ complexity, dominating the ProSparsity processing overhead, so we can safely ignore the rest of the two terms. 
Using ProSparsity, we can reduce $\Delta S \times m \times k \times n$ floating-point additions, where $\Delta S$ denotes the increase in sparsity introduced by ProSparsity.
Our experiments indicate that a floating-point addition incurs $45 \times$ the hardware overhead of a single TCAM bitwise operation. 
ProSparsity processing is more efficient than direct computation when the benefit-cost ratio exceeds one:
$$\frac{\Delta S \times m \times k \times n \times 45}{m^2 \times k \times 1} > 1,$$
where we can get the threshold as $\Delta S = 4.4\%$.
That means that \name can benefit when the sparsity increase is larger than $4.4\%$.
Given our current tile size ($m=256, k=16, n=128$) and average sparsity change in \Sec{sec:sparsity_comparison} ($\Delta S = 13.35\%$), the benefit-cost ratio reach $3.0 \times$.
It confirms that our approach yields a favorable trade-off between processing overhead and computation reduction, thus validating the efficiency of our proposed method.
}
\section{\revision{Discussion and Related Works}}

\subsection{
\revision{Architecture Scalability}}
\revision{
\name has the potential for good scalability. 
Both intra-PPU and inter-PPU parallelism could be leveraged to scale up the Prosperity architecture. 

\textbf{Intra-PPU.}
Intra-PPU parallelism could be achieved by simultaneously issuing multiple tasks (i.e., nodes in the ProSparsity tree) to the processor. 
There are no dependency for nodes in the same level.
Therefore, we can scale to parallel process multiple nodes simultaneously.
}

\revision{
\textbf{Inter-PPU.}
}
\revision{Furthermore, Prosperity's scalability can be extended by exploring inter-PPU parallelism through deploying multiple PPUs within the system. 
Each PPU can process one tile of the spiking GeMM at a time, and large models can be divided into multiple tiles.
}

\subsection{SNN Accelerators}
Besides the work discussed in \Sec{sec:snn_accelerator}, there have been extensive work utilizing the bit sparsity to efficiently accelerate SNNs~\cite{narayanan2020spinalflow, liu2022sato, yin2022sata, yin2024loas, khodamoradi2021s2n2}. SpinalFlow~\cite{narayanan2020spinalflow} is the first architecture designed for SNNs. It target on temporal coded SNNs to utilize BitSparsity and proposes time batching to reduce data movement.
\revision{SATO~\cite{liu2022sato} utilize bucket-sort to evenly distribute the task of sparse addition to multiple PEs.
SATA~\cite{yin2022sata} is a sparsity aware SNN accelerator that build upon systolic array, with a special support for SNN training.}

\revision{Recent works have introduced DNN techniques like quantization and pruning to SNNs for more efficient accelerators. MINT~\cite{yin2024mint}, built upon SATA~\cite{yin2022sata}, incorporates weight and membrane potential quantization to reduce memory footprint. LoAS~\cite{yin2024loas}, the accelerator considering both spike bit sparsity and pruned weight sparsity, proposes a parallel dataflow for dual-side sparsity processing. Our ProSparsity introduces a novel sparsity paradigm in SNN activation, orthogonal to weight quantization and pruning, indicating potential for future integration with these techniques.}

\subsection{Efficient DNN}
\textbf{Quantitation and Sparsity.}
Compression techniques such as quantization~\cite{han2015deep, yao2020zeroquant, guo2022squant, lin2023awq, frantar2023gptq, dettmers2022llm, guo2024survey} and sparsity~\cite{han2015deep, wen2016learning, guo2020accelerating, NIPS1989_6c9882bb, zhang_h_2o_2023,  guo2024accelerating} are widely employed in DNN accelerators~\cite{guo2022ant, zadeh2022mokey, zadeh2020gobo, guo2023olive, wang2021dual, zhang2024dstc, han2016eie, chen2018regan, yan2019rram} to enhance both computational and memory efficiency for DNN models. 
Similar to LoAS~\cite{yin2024loas} or MINT~\cite{yin2024mint}, these methods leverage pruning or quantization to compress weight tensors, thereby enabling further acceleration in SNN. 
Notably, \name is fundamentally orthogonal to most quantization and pruning (sparsity) techniques, allowing it to remain compatible with quantized and sparse weights for continued computational optimization. 
This synergy opens up new avenues for future research.

\textbf{Redundancy Removal.} Redundancy removal methods have been applied in Graph Neural Networks~\cite{kipf2016semi,fu2024desco} (GNNs) accelerators~\cite{geng2021gcn,chen2022regnn}, categorized into offline and online approaches. Offline methods~\cite{jia2020redundancy} perform global graph traversal to optimize redundancy elimination but incur heavy overhead, which is acceptable for static GNN graphs as it is a one-time process. In contrast, online methods~\cite{geng2021gcn,chen2021rubik} use heuristic graph reordering to identify and eliminate redundancy locally, achieving efficiency in hardware but with less redundancy removed than offline methods.

However, these methods are GNN-specific and unsuitable for SNNs due to two main reasons. First, GNN methods leverage the presence of communities~\cite{girvan2002community} in real-world graphs~\cite{broder2000graph,redner1998popular,zhang2024qplacer}, which SNN sparsity patterns lack, exhibiting random distributions due to dynamic input activation. Second, GNNs typically handle highly sparse graphs ($\geq$99.99\% sparsity)\cite{geng2021gcn,backstrom2006group,dai2022dimmining}, whereas SNNs have lower sparsity (60\%–90\%)\cite{li2021differentiable,lv2023spikebert,zhou2022spikformer}, making redundancy identification fundamentally different.

\section{Conclusion}
In conclusion, this study introduces ProSparsity, a groundbreaking sparsity paradigm for SNNs that significantly enhances computational efficiency beyond traditional bit sparsity. 
By leveraging combinatorial similarities within spike matrices, ProSparsity enables substantial reductions in computations and energy consumption. 
Our proposed architecture, \name, effectively addresses the spatial and temporal challenges of implementing ProSparsity, achieving linear time complexity and overhead-free sparse processing.
The results demonstrate \name's superiority over state-of-the-art SNN accelerators, with average speedups of 7.4$\times$ and 1.8$\times$ compared to PTB and A100 GPU, respectively. 
Moreover, \name achieves remarkable energy efficiency improvements of 8.0$\times$ and 193$\times$ over these benchmarks. 
These advancements not only push the boundaries of SNN acceleration but also open new avenues for efficient AI processing across diverse applications.

\section*{Acknowledgements}
This work was supported in part by NSF grants 2328805 and 2112562, as well as Duke University's internal grant, the Beyond the Horizon Initiative. The authors thank the anonymous reviewers for their constructive feedback, which helped improve this work. We also extend our gratitude to Bowen Duan, Tergel Molom-Ochir, Jonathan Ku, Dr. Ziru Li, and Dr. Yangjie Zhou for their technical support and insightful discussions.
%
%
%
%
%

\newpage
\appendix

\section{Artifact Appendix}

\subsection{Abstract}
Our artifact is the simulator for \name, which evaluates the runtime performance and energy.

We evaluate \name on SNN models based on CNN model architecture or transformer architecture for image recognition and natural language processing tasks. 
For spiking CNN models, we have VGG16 and ResNet18. 
For spiking transformers, they include Spikformer, SDT, SpikeBERT, and SpikingBERT. 
These models are tested on eight dataset from NLP to image recognition, including SST-2, SST-5, MR, QQL, MNLI, CIFAR10, CIFAR100, CIFAR10-DVS.
We provide the sparse activation matrix of models on these datasets, it would be used for architecture performance simulation.
Our simulator evaluates the performance of \name and baseline accelerators according to the provided sparse matrices.
We run all the experiments on a Ubuntu server with an NVIDIA A100 GPU.

\subsection{Artifact check-list (meta-information)}

{\small
\begin{itemize}
  \item {\bf Compilation:} NVCC 12.1, GCC/G++ 11.3.0.
  \item {\bf Model:} Spiking VGG, Spiking ResNet, Spikformer, Spike-Driven-Transformer, SpikeBERT, SpikingBERT.
  \item {\bf Data set:} GLUE dataset, SST-5, MR, CIFAR10, CIFAR100, CIFAR10-DVS.
  \item {\bf Run-time environment:} Ubuntu 22.04.2 LTS, CUDA 12.1, PyTorch 2.3.0.
  \item {\bf Hardware:} A server with an x86\_64 processor, an NVIDIA A100 GPU.
  \item {\bf Output:} Architecture runtime performance, energy, matrix density.
  \item {\bf How much disk space required (approximately)?:} 8GB.
  \item {\bf How much time is needed to prepare workflow (approximately)?:} It takes about 30 minutes to prepare the environment.
  \item {\bf How much time is needed to complete experiments (approximately)?:} It takes about 2 hours to run all of the experiments including architecture simulation, design space analysis, matrix sparsity analysis.
  \item {\bf Publicly available?:} Our framework is publicly available on
GitHub \url{https://github.com/dubcyfor3/Prosperity}
  \item {\bf Code licenses (if publicly available)?:} MIT license.
  \item {\bf Data licenses (if publicly available)?:} The datasets are publicly available through their original licensing terms.
  \item {\bf Archived (provide DOI)?:} \url{https://doi.org/10.5281/zenodo.14350604}
\end{itemize}
}


\subsection{Description}

\subsubsection{How to access} We archive the source code at \url{https://doi.org/10.5281/zenodo.14350759}. We recommend you access
our GitHub repository: \url{https://github.com/dubcyfor3/Prosperity} for the latest version.

\subsubsection{Hardware dependencies} A server equipped with an NVIDIA GPU.

\subsubsection{Software dependencies} The experiments rely on the following software components.
\begin{itemize}
    \item Ubuntu 22.04.2 LTS
    \item Python 3.10
    \item PyTorch 2.3.0
    \item Anaconda 24.1.2
    \item G++ 11.3.0
    \item CUDA 12.1
    \item CACTI 7.0
\end{itemize}

\subsubsection{Data sets and models} The computer vision model including spiking VGG-16, ResNet-18, and Spikformer, Spike-Driven-Transformer models are run on image classification dataset CIFAR10, CIFAR100, CIFAR10-DVS. NLP models including SpikeBERT, SpikingBERT are run on NLP dataset GLUE (SST-2, MNLI, QQP), SST-5, MR.

\subsection{Installation}
We have well-documented README file to detail the
installation instruction for each experiment at \url{https://github.com/dubcyfor3/Prosperity}.

\subsection{Experiment workflow}

The README file also specifies the detailed experiment workflow to get the results reported in paper.
\subsection{Evaluation and expected results}

Our artifact evaluate the runtime performance of our proposed Prosperity and baselines. It also include the design space exploration experiment, sparsity analysis, and buffer area evaluation.

\subsection{Methodology}

Submission, reviewing and badging methodology:

\begin{itemize}
  \item \url{https://www.acm.org/publications/policies/artifact-review-and-badging-current}
  \item \url{https://cTuning.org/ae}
\end{itemize}


\bibliographystyle{IEEEtranS}
\bibliography{refs}

\end{document}